\documentclass[12pt, a4paper]{report}

\usepackage{iftese} 
\usepackage{axodraw}
\usepackage{latexsym}
\usepackage{graphics}

\renewcommand{\baselinestretch}{1.2} 
\newfont{\got}{cmfrak at 12pt}  
\newfont{\cmsy}{cmsy10 at 12pt} 
\newfont{\cmmi}{rsfs10 at 12pt} 
\newfont{\cmss}{cmss10 at 12pt} 

\newcommand \dslash {\partial\!\!\!/}
\newcommand \del {\partial}
\newcommand \tr {T\!r}
\newcommand \Z {Z\!\!\!Z}

\newcommand{\al} {\ensuremath{\alpha}}
\newcommand{\be} {\ensuremath{\beta}}
\newcommand{\lrarrow}[1]{\stackrel{\leftrightarrow}{#1}}
\newcommand \beq {\begin{equation}}
\newcommand \eeq {\end{equation}}
\newcommand \DS {\displaystyle}
\newcommand \ST {\scriptstyle}
\newcommand \sig {\ensuremath{\sigma}~}

\newcommand \btheta {\bar{\vartheta}}
\newcommand \Q {\mbox{\cmmi Q}}
\newcommand \D {\mbox{\cmmi D}}
\newcommand \DB {\bar{\mbox{\cmmi D}}}
\newcommand \V {\mbox{\cmmi V}}

\newcommand \M {\ensuremath{\mbox{\cmsy M}\,}}
\newcommand \N {\ensuremath{\mbox{\cmsy N}\,}}

\newcommand \RR {\ensuremath{R\!\!\!\!R}}
\newcommand \s {\ensuremath{{\cal S}}}

\newcommand \X {\ensuremath{X\!\!\!\!X}}
\newcommand \CC {\ensuremath{C\!\!\!\!C}}

\newcommand {\gs}{G\!\!\!\!G}

\newcommand {\mb}{\mathbf}
\newcommand {\mr}{\mathrm}
\newcommand {\Tor}[2]{\ensuremath{\mr{Tor}(\mb{#1},\mb{#2})}}
\newcommand {\DD}[2]{\ensuremath{{\mr{D}_{\mb{#1}}}\mb{#2}}}
\newcommand {\GG}[2]{\ensuremath{{\mr{g}(\mb{#1},\mb{#2})}}}
\newcommand {\lie}[2]{\ensuremath{{\mr{\pounds}_{\mb{#1}}}\mb{#2}}}

\newcommand {\meio}{\ensuremath{\frac{1}{2}}}

\newcommand \bolA {
\begin{picture}(40,30)(0,0)
\Line(0,14.5)(20,14.5)
\Line(0,15.5)(20,15.5)
\Vertex(20,15){1}
\ArrowArc(30,15)(10,0,180)
\ArrowArc(30,15)(10,180,360)
\Text(2,17)[bl]{$\scriptstyle A$}
\end{picture}}

\newcommand \bolB {
\begin{picture}(40,30)(0,0)
\Line(0,14.5)(20,14.5)
\Line(0,15.5)(20,15.5)
\Vertex(20,15){1}
\ArrowArc(30,15)(10,0,180)
\ArrowArc(30,15)(10,180,360)
\Text(2,17)[bl]{$\scriptstyle B$}
\end{picture}}

\newcommand \diagAA {
\begin{picture}(60,30)(0,0)
\Line(0,14.5)(20,14.5)
\Line(0,15.5)(20,15.5)
\Vertex(20,15){1}
\Vertex(40,15){1}
\ArrowArc(30,15)(10,0,180)
\ArrowArc(30,15)(10,180,360)
\Line(40,14.5)(60,14.5)
\Line(40,15.5)(60,15.5)
\Text(2,17)[bl]{$\scriptstyle A$}
\Text(58,17)[br]{$\scriptstyle A$}
\end{picture}}

\begin{document}

\def\baselinestretch{1.2}
\hoffset=-1.0 true cm
\voffset=-2 true cm
\topmargin=1.0cm
\thispagestyle{empty}

\thicklines
\begin{picture}(370,60)(0,0)
\setlength{\unitlength}{1pt}
\put(40,53){\line(2,3){15}}
\put(40,53){\line(5,6){19}}
\put(40,53){\line(1,1){27}}
\put(40,53){\line(6,5){33}}
\put(40,53){\line(3,2){25}}
\put(40,53){\line(2,1){19}}
\put(40,53){\line(5,-6){17}}
\put(40,53){\line(1,-1){22}}
\put(40,53){\line(6,-5){30}}
\put(40,53){\line(3,-2){22}}
\put(40,53){\line(-2,1){15}}
\put(40,53){\line(-3,1){23}}
\put(40,53){\line(-4,1){26}}
\put(40,53){\line(-6,1){36}}
\put(40,53){\line(-1,0){40}}
\put(40,53){\line(-6,-1){32}}
\put(40,53){\line(-3,-1){20}}
\put(40,53){\line(-2,-1){10}}
\put(75,45){\Huge \bf IFT}
\put(180,56){\small \bf Instituto de F\'\i sica Te\'orica}
\put(165,42){\small \bf Universidade Estadual Paulista} 
\put(-25,2){\line(1,0){433}}
\put(-25,-2){\line(1,0){433}}
\end{picture}  


\vskip .3cm
\noindent
{DISSERTA\c C\~AO DE MESTRADO}
\hfill    IFT--D.002/99\\


\vspace{3cm}
\begin{center}
{\large \bf Explorando o Modelo $\sigma$ Supersim\'etrico}

\vspace{1.2cm}
Breno Cesar de Oliveira Imbiriba.
\end{center}

\vskip 3cm
\hfill Orientador
\vskip 0.4cm
\hfill {\em Prof. Dr. Maria Cristina Batoni Abdalla.}
\vskip 0.4cm
\hfill Co-orientador
\vskip 0.4cm
\hfill {\em Prof. Dr. Nathan J. Berkovits.}
\vskip 4cm
\vfill
\begin{center}
Junho de 1999
\end{center}

\newpage

\pagenumbering{roman}

\begin{center}
{\Large \bf Agradecimentos}
\end{center}
\vskip 2.0cm
Agrade\c co muito a todos meus colegas e amigos do IFT, especialmente a:
{\cmss Alfonso Zerwekh, Brenno Vallilo, Cristiane Rold\~ao, \'Erica Emilia Leite, Evelise Gausmann, Jorge Casti\~neiras, Ricardo Bent\'{\i}n, Vanessa Andrade, Victo Filho, Dr. Maria Caballero Tijero, Dr. Paulo S\'ergio e Dr. Marcelo Leite,} com os quais tive discuss\~oes relevantes para o compreendimento deste assunto. Tamb\'em agrade\c co a {\cmss Carlos Pires, Cristiano Santos, Daniel Vanzella, Daniel Nedel, Jose Acosta, Luis Crispino e Marcel Ferreira}. Em especial agrade\c co ao {\cmss Prof. Dr. George A. Matsas}.

Agrade\c co enormemente a {\cmss Alexandre Guimar\~aes Rodrigues} e {\cmss Dr. Jos\'e M. F. Bassalo}, pelo apoio acad\^emico e pessoal ao longo de minha estadia em S\~ao Paulo. 

Agrade\c co aos meus familiares por inestim\'avel apoio durante os estudos deste mestrado, em especial a {\cmss Dr. Nazar\'e Imbiriba Mitschein}, que sem seu apoio nada poderia ser poss\'{\i}vel, {\cmss Dr. Thomas Mitschein, Dr. Miguel Imbiriba, Miguel Imbiriba Jr, Tales} e {\cmss Lucas Imbiriba, Luisa Mitschein,} \`a fam\'{\i}lia {\cmss Titan, Orlandina} e {\cmss Alirio Oliveira}, e {\cmss Octavio C. Dourado Jr}. 

Agrade\c co tamb\'em aos meus amigos que mesmo distantes n\~ao me deixaram desanimar, em especial a: {\cmss Waleska Barros, Andr\'e Siilva, Paulo Daniel, Ana Carina, Andrey Miralha,} ao grande {\cmss Kor, o Mal\'{\i}gno}, a {\cmss Garnedon, Warlock, Jaufre Rudel de Blaia, Yantl, Yorel de Daggerford,} a {\cmss Leugim, o a\c cougueiro de Tempus}, aos {\cmss ``Lordes da Escurid\~ao''} e ao {\cmss Exercitus Tenebrarum}.  

Agrade\c co de todo o cora\c c\~ao pelo apoio e ajuda nestes dois anos \`a {\cmss Silvia Ataide da Silva.}

Agrade\c co muit\'{\i}ssimo \`a {\cmss Prof. Maria Cristina B. Abdalla}, pela orienta\c c\~ao nas quest\~oes complicadas da F\'{\i}sica e pela inestim\'avel ajuda na feitura deste trabalho, e ao {\cmss Prof. Nathan J. Berkovits}, que al\'em de orientar, magnificamente bem, se disp\^os a tocar contra-baixo conosco.   

Agrade\c co \`a {\cmss FAPESP}, sem a qual este trabalho seria enormemente dificultado.

\newpage

\begin{center}
{\Large \bf Resumo}
\end{center}
\vskip 2.0cm


Este trabalho visa apresentar o modelo sigma n\~ao linear de maneira simples e direta. Em sua primeira parte mostramos o modelo bos\^onico e o termo de Wess-Zumino-Witten sobre o qual tecemos coment\'arios sobre seu car\'ater topol\'ogico e sua associa\c c\~ao com a tor\c c\~ao. Posteriormente mostramos que para cancelar a anomalia conforme qu\^antica, o modelo passa a obedecer as Equa\c c\~oes de Einstein. Introduzimos um curto cap\'{\i}tulo sobre supersimetria a fim de auxiliar na exposi\c c\~ao do modelo supersim\'etrico. No \'ultimo cap\'{\i}tulo apresentamos o modelo supersim\'etrico e suas equa\c c\~oes de movimento. Finalmente apresentamos o caso com duas supersimetrias, introduzindo os campos quirais e quirais torcidos e explicitamos o modelo para o caso espec\'{\i}fico $SU(2)\otimes U(1)$.


\vskip 1.0cm
\noindent
{\bf Palavras Chaves}: Modelo Sigma n\~ao Linear; Modelo Sigma Supersim\'etrico; Cordas; Variedades de Grupos.  
\vskip 0.5cm
\noindent
{\bf \'Areas do conhecimento}: Teoria de Campos (1050300-5); F\'{\i}sica Matem\'atica (1010401-1)

\newpage

\begin{center}
{\Large \bf Abstract}
\end{center}
\vskip 2.0cm

The purpose of this work is to present some basic concepts about the non-linear sigma model in a simple and direct way. We start with showing the bosonic model and the Wess-Zumino-Witten term, making some comments about its topological nature, and its association with the torsion. It is also shown that to cancel the quantum conformal anomaly the model should obey the Einstein equations. We provide a quick introduction about supersymmetry in chapter 2 to help the understanding the supersymmetric extension of the model. In the last chapter we present the supersymmetric model and its equations of motion. Finaly we work-out the two-supersymmetry case, introducing the chiral as well as the twisted chiral fields, expliciting the very specific $SU(2)\otimes U(1)$ case.

\vskip 1.0cm
\noindent
{\bf Key Words}: Non-linear Sigma Model; Supersimmetric Sigma Model; Stings; Group Manifolds.  
\vskip 0.5cm

\vfill \eject

\pagenumbering{roman}
\tableofcontents


\pagenumbering{arabic}


\chapter*{Introdu\c{c}\~ao}
\addcontentsline{toc}{chapter}{Introdu\c{c}\~ao}
\markright{Introdu\c{c}\~ao}

Nesta disserta\c{c}\~ao buscamos introduzir de maneira simples e direta o modelo sigma n\~ao linear e suas formula\c{c}\~oes mais conhecidas. 
Este trabalho divide-se basicamente em duas partes, uma inicial onde introduzimos o modelo sigma bos\^onico, e uma segunda parte na qual tratamos de aspectos supersim\'etricos do modelo e conclu\'{\i}mos com um exemplo N=2. H\'a uma parte intermedi\'aria, curta, que tenta introduzir alguns fundamentos de supersimetria. 

No primeiro cap\'{\i}tulo, iniciamos com a abordagem geom\'etrica do modelo sigma. Este \'e introduzido atrav\'es de um mapeamento entre duas variedades (espa\c{c}o alvo e espa\c{c}o-tempo) e com o aux\'{\i}lio de um tensor do espa\c{c}o alvo, sim\'etrico, de posto 2 definido como a m\'etrica deste espa\c{c}o. Este tratamento evidencia o car\'ater geom\'etrico do caso bi-dimensional, onde a a\c{c}\~ao da corda bos\^onica mostra-se como sendo a \'area do espa\c{c}o-tempo (folha-mundo da corda) medida pelo espa\c{c}o-alvo. Seguindo esta linha introduz-se o termo de Wess-Zumino-Witten (WZW) como um tensor antissim\'etrico de posto $n$. Atrav\'es do teorema de Stokes, este termo pode ser estendido para um tensor antissim\'etrico de posto $n+1$, permitindo que termo de WZW seja definido em uma extens\~ao $(n+1)$-dimensional do espa\c{c}o-tempo. Sua exist\^encia requer algumas propriedades topol\'ogicas do espa\c{c}o alvo, a saber, que este seja simplesmente conexo e compacto. Este termo revela uma quantiza\c{c}\~ao da constante de acoplamento que \'e conhecida em teoria de cordas como {\it winding number}.

Seguindo, introduzimos o modelo sigma sobre variedades de grupos. Esta abordagem talvez seja a mais usual e foi a que originou o modelo. Esta apresenta\c{c}\~ao do modelo \'e mais restritiva e mais sim\'etrica uma vez que variedades de grupos tem alta simetria, ou seja, uma quantidade grande de vetores de Killing. O termo de Wess-Zumino-Witten \'e introduzido e suas restri\c{c}\~oes topol\'ogicas s\~ao apresentadas. A equival\^encia entre as duas formula\c{c}\~oes do modelo \'e mostrada. Por\'em no caso de duas dimens\~oes, para haver uma associa\c{c}\~ao mais completa entre as formula\c{c}\~oes, uma an\'alise da tor\c{c}\~ao de variedades de grupo se faz necess\'aria, uma vez que o tensor antissim\'etrico introduzido na formula\c{c}\~ao geom\'etrica pode ser associado com a tor\c{c}\~ao do espa\c{c}o alvo. 

Este cap\'{\i}tulo, encerra-se com um c\'alculo qu\^antico. No caso do espa\c{c}o-tempo ser bi-dimensional (cordas), a a\c{c}\~ao apresenta invari\^ancia conforme no n\'{\i}vel cl\'assico. Para que isto se mantenha quanticamente uma restri\c{c}\~ao deve ser observada; que o espa\c{c}o alvo obede\c{c}a a equa\c{c}\~ao de Einstein. Em cordas, o espa\c{c}o alvo \'e o espa\c{c}o-tempo f\'{\i}sico, ent\~ao a condi\c c\~ao sobre a anomalia conforme implica que, no n\'{\i}vel de \'arvore (folha mundo de genus zero), a Relatividade Geral \'e preservada.

No cap\'{\i}tulo dois, introduzimos alguns conceitos de supersimetria. A \'algebra de supersimetria \'e introduzida de maneira gradativa e apenas depois a sua formaliza\c{c}\~ao \'e feita. Os geradores de supersimetria s\~ao definidos em analogia com o gerador de transla\c{c}\~ao, o momento. Ap\'os, as derivadas covariantes e os supercampos s\~ao introduzidos. Um exemplo de supersimetria no cone de luz em duas dimens\~oes \'e apresentado uma vez que este ser\'a a representa\c{c}\~ao que utilizaremos na parte final. Este cap\'{\i}tulo termina com uma breve exposi\c{c}\~ao de m\'ultiplas supersimetrias j\'a que parte deste assunto ser\'a explicado no momento oportuno.

O terceiro cap\'{\i}tulo aplica as id\'eias dos dois primeiros para introduzir o modelo sigma n\~ao linear supersim\'etrico N=1 em duas dimens\~oes. Suas equa\c{c}\~oes de movimento s\~ao apresentadas, tanto na forma manifestamente supersim\'etrica como em componentes.
 ~\'E importante observar neste ponto que a transi\c{c}\~ao do modelo N=0 para N=1 n\~ao exige nenhuma restri\c{c}\~ao das variedades do espa\c{c}o-tempo ou espa\c{c}o-alvo. 

O caso N=2 apresenta mais sutilezas que o anterior. Inicialmente, por raz\~oes dimensionais, n\~ao podemos definir o modelo sigma com supersimetria N=2 manifesta da mesma forma que nos casos N=0 e N=1. Na verdade, a \'unica maneira de se definir uma a\c{c}\~ao com duas supersimetrias manifestas, em duas dimens\~oes, \'e atrav\'es de uma lagrangeana escalar. Isto faz com que a \'unica fonte de din\^amica do modelo se apresente por meios de v\'{\i}nculos que os supercampos venham a ter. 

Apresentamos dois tipos de supercampos N=2, que cont\^em v\'{\i}nculos sobre suas componentes N=1, limitando seus ``movimentos''. Estes s\~ao os campos complexos quiral e o quiral torcido. Os v\'{\i}nculos que suas componentes sofrem permitem que escrevamos cada um desses campos em termos de apenas uma de suas componentes. Desta forma podemos reduzir o modelo de N=2 para N=1. Um fato importante a ressaltar \'e que neste ponto o espa\c{c}o alvo j\'a n\~ao pode ser mais arbitr\'ario. Os campos complexos introduzidos, podem ser vistos como coordenadas de uma variedade complexa de K\"ahler. 

A se\c{c}\~ao final apresenta o modelo sigma N=2 sobre a variedade de grupo $SU(2)\otimes U(1)$, que \'e K\"ahler, uma vez que tem grupo de holonomia $U(2)$. Este grupo permite que escrevamos o modelo sigma de maneira supersim\'etrica atrav\'es de campos quirais e quirais torcidos. Neste exemplo mostramos explicitamente como o potencial de K\"ahler nos define a m\'etrica do grupo, logo o modelo principal. O termo de WZW n\~ao pode ser introduzido como algo extra, ele est\'a completamente contido no potencial e se caracterizar\'a como a tor\c{c}\~ao desta variedade. Calculando os termos antissim\'etricos da a\c{c}\~ao e comparando com a defini\c{c}\~ao bos\^onica do termo de WZW, vemos que a tor\c{c}\~ao deste modelo \'e proporcional \`aquela definida nos outros casos com menos supersimetrias.

Esperamos que esta monografia tenha cumprido com o seu papel de explorar um assunto n\~ao trivial e ao mesmo tempo interessante, que certamente contribuiu muito para o aprendizado do autor sobre o assunto e que possa futuramente servir para outros alunos do Instituto de F\'{\i}sica Te\'orica.


\chapter{O Modelo \sig N=0}\label{cap.sigmabos}

\section{Introdu\c{c}\~ao}


Modelos sigma n\~ao lineares t\^em sido amplamente estudados tanto no contexto de mec\^anica estat\'{\i}stica como em teoria qu\^antica de campos. Historicamente foram introduzidos por Schwinger \cite{Schwinger} e usados em teoria de campos por Gell-Mann e Levy nos anos sessenta para descrever a fenomenologia dos p\'{\i}ons \cite{GellMann}. Originalmente introduzido em quatro dimens\~oes e utilizado para descrever a fenomenologia de PCAC (corrente axial parcialmente conservada), n\~ao tinha inicialmente car\'ater geom\'etrico. No entanto, um dos percal\c cos do modelo foi a previs\~ao de uma part\'{\i}cula (a part\'{\i}cula \sig) que nunca foi observada. Uma maneira de retirar esta part\'{\i}cula da teoria seria implementar a simetria quiral $SU(2)$ de forma n\~ao linear, impondo um v\'{\i}nculo. O modelo passa ent\~ao a ser descrito pela geometria da variedade alvo onde \'e definido, ou seja a esfera $S^3$. Infelizmente o modelo em quatro dimens\~oes n\~ao \'e renormaliz\'avel \cite{Honerkamp}. Junte-se a esse fato o grande sucesso das teorias de gauge n\~ao-Abelianas e fica claro porque o modelo sigma acabou sendo colocado em segundo plano.

J\'a em duas dimens\~oes o modelo sigma n\~ao linear aparece como um laborat\'orio promissor no estudo de diversos aspectos de teoria qu\^antica de campos. Tais modelos, tendo origem geom\'etrica, aproximam-se das teorias de gauge n\~ao-Abelianas. O sucesso do modelo com simetria $O(N)$ que mostrou-se ser integr\'avel, ter matriz $S$ exata, exist\^encia de ``gap'' de massa, inspirou generaliza\c c\~oes procurando preservar o fato do modelo ser definido em um espa\c co de base geom\'etrica, ou seja sobre uma variedade diferencial.

A primeira generaliza\c c\~ao \'e proposta por Eichenherr \cite{Eichenherr} em 1978 onde o $CP^{N-1}$ \'e proposto que \'e a generaliza\c c\~ao para o grupo $SU(N)$. No in\'{\i}cio da d\'ecada de 80, demosntra-se que os modelos sigma em suas dimens\~oes s\~ao renormalizaveis \cite{fridan}, e estes  voltam ao cen\'ario dentro do contexto de \'algebras de correntes e bosoniza\c c\~ao n\~ao-Abeliana onde neste momento o termo de Wess-Zumino \'e agregado ao modelo sigma por Witten \cite{Witten2, Witten1} em 1983.
Em 84-85 o modelo de Wess-Zumino \'e  supersimetrizado \cite{susy} e seus aspectos come\c cam a ser estudados. 

Um outro cen\'ario para o modelo sigma \'e a teoria de cordas. Esta foi inicialmente introduzida por Nambu-Goto\cite{nambu}, por\'em, devido a n\~ao linearidade da a\c c\~ao, esta foi substituida pela a\c c\~ao de Polyakov\cite{pol}, equivalente e bem mais trat\'avel. Desde ent\~ao, o modelo sigma passou a ser considerado como fundamental para a teoria de cordas e supercordas.

\section{Considera\c{c}\~oes Gerais.}

Antes de definir a a\c{c}\~ao do modelo \sig, precisamos introduzir dois conceitos:
\begin{itemize}
\item{Espa\c{c}o-Alvo}: Variedade m-dimensional \M munida de uma m\'etrica g, $(\M,\mr{g})$, parametrizada pelas coordenadas $\phi^i, i=0,...,m-1$.
\item{Espa\c{c}o-Tempo}: Variedade n-dimensional \N munida de uma m\'etrica $\gamma$, $(\N,\gamma)$, parametrizada pelas coordenadas $x^\mu, \mu=1,...,n$.
\end{itemize}

Consideremos uma imers\~ao\footnote{Uma imers\~ao define-se por ser localmente bijetora, ou seja, leva vizinhan\c{c}as de um ponto $x\in \N$ \`a vizinhan\c{c}as de $\phi(x)\in \M$ e sua inversa, restrita \`a vizinhan\c{c}a de $\phi(x)$, leva pontos $\phi\in \M$ a pontos $x\in \N$ contidos na vizinhan\c{c}a de $x$ (mapa sobrejetor). E se todos os pontos de $\N$ forem mapeados a pontos de $\M$ (mapa injetor).} de \N em \M definida pelo mapa $\phi^i=\phi^i(x)$. Ou seja teremos que \N define um sub-espa\c{c}o em \M atrav\'es de $\phi$.

O modelo \sig n\~ao-linear principal \'e definido como\footnote{Na teoria de cordas, onde $n=2$, o espa\c{c}o-tempo \'e a folha-mundo e o espa\c{c}o-alvo \'e o espa\c{c}o f\'{\i}sico.}:


\begin{equation}\label{eq.acao}
{\cal S}=\mu^{l}\int d^nx\sqrt{\gamma}\gamma^{\mu\nu}\mr{g}_{ij}\del_\mu\phi^i\del_\nu\phi^j,
\end{equation}
onde, $\gamma=\gamma^{\mu\nu}(x)$, $ \mr{g}=\mr{g}_{ij}(\phi)$. A constante $\mu$ tem dimens\~ao de massa.\footnote{Basta observar que se considerarmos os campos e a a\c{c}\~ao adimensionais, com $\hbar=c=1$, teremos que $\mu^l$ assume dimens\~ao de $[massa]^{l}$, com $l=n-2$. 
} 

Pensando em uma imers\~ao, podemos considerar  $\del_\mu\phi^i\del_\nu\phi^j\mr{g}_{ij}$ como uma transforma\c{c}\~ao de coordenadas de \M para \N. Definamos a m\'etrica induzida sobre a variedade \N como sendo:
\beq
h_{\mu\nu}=\del_\mu\phi^i\del_\nu\phi^j\mr{g}_{ij}.
\eeq

 
As equa\c{c}\~oes de movimento podem ser extra\'{\i}das via varia\c{c}\~ao da a\c{c}\~ao pelos campos $\phi$, e s\~ao da forma:
\beq \label{eq.sigmamov}
\del_\mu(\del^\mu\phi^k)+\Gamma^k_{ij}(\del_\mu\phi^i)(\del^\mu\phi^j)=0,
\eeq
onde $\Gamma^k_{ij}$ \'e o s\'{\i}mbolo de Christoffel da variedade \M definido como: $\Gamma^k_{ij}=\frac{1}{2}\mr{g}^{kr}(\mr{g}_{ir,j}+\mr{g}_{jr,i}-\mr{g}_{ij,r})$. 
Para compreender melhor o significado desta a\c{c}\~ao, observemos que a equa\c{c}\~ao de uma geod\'esica $\lambda(t)$ de \M \'e:
\beq
d^2_t\phi^a+\Gamma^a_{bc}d_t\phi^bd_t\phi^c=0.
\eeq
As equa\c{c}\~oes de movimento do modelo \sig s\~ao, na verdade, uma generaliza\c{c}\~ao do conceito de geod\'esicas, no sentido de que n\~ao temos mais uma curva $\lambda(t)$ parametrizada por um par\^ametro e sim um sub-espa\c{c}o de \M, $\phi(\N)$, que \'e parametrizado pelas vari\'aveis $x^\mu$\footnote{Por exemplo, se $n=2$ e $\Gamma=0$, teremos $\del_\mu\del^\mu\phi^l=0$. N\~ao \'e dif\'{\i}cil mostrar que a solu\c{c}\~ao \'e uma superf\'{\i}cie plana}. 

\vskip 0.5cm

Outro ponto relevante \'e calcular as equa\c c\~oes de movimento da m\'etrica $\gamma$. Veremos que estas equa\c c\~oes s\'o ter\~ao solu\c c\~oes quando o espa\c co-tempo tiver duas dimens\~oes.

Calculemos ent\~ao a varia\c c\~ao da a\c c\~ao com respeito \`a $\gamma$,

\beq\label{eq.variacao}
\delta_\gamma \s=\int_{\cal N} d^nx\sqrt{\gamma}\left(\frac{1}{2}\gamma_{\mu\nu}\gamma^{\al\be}h_{\al\be}+h_{\mu\nu}\right)\delta \gamma^{\mu\nu}=0.
\eeq
Desta maneira teremos que
\beq
h_{\mu\nu}=-\frac{1}{2}\gamma_{\mu\nu}(\gamma^{\al\be}h_{\al\be}).
\eeq
Dividindo pelo determinante de $h_{\mu\nu}$, teremos que:
\beq
\frac{h_{\mu\nu}}{\sqrt{h}}=\frac{\gamma_{\mu\nu}}{\sqrt{h}},
\eeq
ou seja, de maneira geral, 
\beq \label{eq.prop}
\gamma_{\mu\nu}(x)=f(x)h_{\mu\nu}(x).
\eeq
Isto significa que m\'etrica $\gamma$ estar\'a relacionada com a m\'etrica induzida $h$ via uma transforma\c c\~ao conforme.
Substituindo a Eq.~(\ref{eq.prop}) em~(\ref{eq.variacao}), teremos ainda que:
\begin{eqnarray}
0=h_{\mu\nu}+\frac{1}{2}\gamma_{\mu\nu}(\gamma^{\al\be}h_{\al\be})&=&h_{\mu\nu}+\frac{1}{2}fh_{\mu\nu}f^{-1}h^{\al\be}h_{\al\be}\nonumber \\
&=&(\frac{n}{2}h_{\mu\nu}+h_{\mu\nu})h^{\mu\nu}=\frac{n-2}{2}n.
\end{eqnarray}
Implicando que $n=2$, ou seja, para a m\'etrica $\gamma$ ser um campo din\^amico e obedecer a Eq.(\ref{eq.prop}), temos que estar em duas dimens\~oes\footnote{Podemos apresentar este resultado de uma outra maneira. Para os observadores em \M, a a\c c\~ao (\ref{eq.acao}) descreve uma hiperf\'{\i}cie $n$-dimensional, \N. O valor de $\cal S$ pode n\~ao ter significado algum para estes observadores, por\'em, consideremos que $\cal S$ seja um n\'umero associado \`a \N. Assim, $\cal S$ deve ser equivalente a algo que dependa apenas de $h$, a m\'etrica induzida. Para que o valor da a\c c\~ao n\~ao varie com mudan\c cas da m\'etrica $\gamma$, que \'e irreal para os habitantes de \M, teremo que fazer $\delta_\gamma {\cal S}=0$, que \'e a Eq. (\ref{eq.variacao}).}. 

Vejamos que atrav\'es de (\ref{eq.prop}), a a\c c\~ao assume a forma
\begin{eqnarray}\label{eq.polnam}
\int d^nx\sqrt{\gamma}\gamma^{\mu\nu}h_{\mu\nu}&=&\int d^nx\sqrt{|fh_{\mu\nu}|}f^{-1}(x)\nonumber \\
&=&\int d^nx\sqrt{h}|f|^{\frac{n}{2}}f^{-1}=\int d^nx\sqrt{h}f^{\frac{n-2}{2}},
\end{eqnarray}
e quando $n=2$ teremos a a\c c\~ao de Nambu-Goto da corda bos\^onica\footnote{Em Eq.~(\ref{eq.polnam}), o termo $d^nx\sqrt{h}$ \'e um infinit\'esimo de hiper-volume, que \'e mensur\'avel a partir de \M. Por outro lado, $f^{\frac{n-2}{2}}$ \'e arbitr\'ario e apenas em $n=2$ ele n\~ao contrinuir\'a, e $\cal S$ ser\'a exatamente a \'area da folha-mundo. Desta forma vemos que a a\c c\~ao original, de Polyakov, reduz-se \`a a\c c\~ao de Nambu-Goto da corda bos\^onica. Tamb\'em podemos ver que a a\c c\~ao $\cal S$ \'e invariante por transforma\c c\~oes conformes de $\gamma$ em duas dimens\~oes.}.

\subsection{Invari\^ancias da a\c{c}\~ao - Termo de WZW.}\label{sec.invariancias}

Podemos enumerar as invari\^ancias da a\c{c}\~ao~(\ref{eq.acao}):
\begin{itemize}

\item \'e invariante por transforma\c{c}\~oes gerais de coordenadas $\phi^i$, no espa\c{c}o $\cal M$.
Para constatar, basta fazer $\phi^i\rightarrow \phi'^i$, transformar $\mr{g}'_{ij}(\phi')$ usando as transforma\c{c}\~oes tensoriais usuais e aplicar a regra da cadeia nos diferenciais, ou seja
\beq
\mr{g'}_{ij}\del_\mu\phi'^i\del_\nu\phi'^j=\mr{g}_{kl}\frac{\del\phi^k}{\del\phi'^i}\frac{\del\phi^l}{\del\phi'^j}\del_\mu\phi'^i\del_\nu\phi'^j=\mr{g}_{kl}\del_\mu\phi^k\del_\nu\phi^l.
\eeq
\item \'e invariante por transforma\c{c}\~oes gerais de coordenadas $x^\mu$, (difeomorfismo) do espa\c{c}o $\cal N$.
Esta invari\^ancia pode ser facilmente verificada uma vez que os campos $\phi^i$ e $\mr{g}_{ij}$ s\~ao campos escalares em $\cal N$, e todos os tensores est\~ao contra\'{\i}dos.

Estas propriedades s\~ao muito importantes por que estas garantem que podemos utilizar o modelo sobre variedades que necessitem de v\'arios mapas, pois passamos de um para outro atrav\'es de uma transforma\c{c}\~ao de coordenadas.
\item {\bf Isometrias} 
Por isometrias tomamos os difeomorfismos que mant\'em as componentes da m\'etrica invariantes. Ou seja, s\~ao transforma\c{c}\~oes que definem as simetrias da variedade.
Para obter as isometrias, consideremos um campo vetorial $k=\frac{d}{dt}$, sobre a variedade \M. Seja $\phi_t$ um difeomorfismo a um par\^ametro na dire\c{c}\~ao de $k$. As transforma\c{c}\~oes $\phi_t$ ser\~ao isometrias se:
\beq
\phi^*_t \mb{g}=\mb{g} \Longrightarrow \lie{k}{g}=0,
\eeq
onde $\lie{k}{}$ \'e a derivada de Lie na dire\c{c}\~ao do vetor $\mb{k}$, $\mb{g}$ \'e a m\'etrica e $\phi^*_t$ \'e o ``pull back'' da transforma\c{c}\~ao. A equa\c{c}\~ao acima pode ser escrita como (por exemplo, veja \cite[pg. 442]{Wald}):
\begin{equation}
\nabla_{(i}k_{j)}(\phi)=0.
\end{equation}

Vetores que satisfazem esta equa\c{c}\~ao s\~ao chamados de {\bf vetores de Killing}, que formam um campo vetorial.
O conjunto de todas as isometrias de uma variedade formam o {\bf grupo de isometrias} {\got{G}}. Assim, por exemplo, se $\cal M$ tem $m$ campos de Killing, ent\~ao podemos ter:
\begin{equation} \label{eq.phitranf}
\delta \phi^i=\Lambda^Ik_I^i(\phi),
\end{equation}
$\Lambda^I$ s\~ao par\^ametros, e $I=1,..,m$.
Os vetores de Killing formam a \'algebra {\got{g}} do grupo de isometrias atrav\'es do comutador:
\beq
[k_i,k_j]=f^l{}_{ij}k_l,
\eeq
onde, $f^l{}_{ij}$ s\~ao as constantes de estrutura do grupo\footnote{Por exemplo, se ${\cal M}=S^3$, a m\'etrica \'e invariante por transforma\c{c}\~oes de O(3), ou seja, as constantes de estrutura do grupo de isometrias seriam as mesmas constantes do grupo O(3).}.




\item {\bf Termos extras}
\end{itemize}

\'E poss\'{\i}vel tamb\'em introduzir alguns novos termos \`a a\c{c}\~ao~(\ref{eq.acao}) os quais n\~ao prejudicar\~ao as invari\^ancias originais. Eles s\~ao:
Um termo potencial $V(\phi)$, que \'e um escalar tanto em $\cal M$ quanto em $\cal N$. 
O termo de Fradkin-Tseytlin: $\Phi(\phi)R^{(\gamma)}(x)$, onde $\Phi$ \'e um escalar (como o $V$), associado ao campo do D\'{\i}laton, e $R^{(\gamma)}$ \'e o escalar de curvatura do espa\c{c}o-tempo $\cal N$.

A adi\c{c}\~ao mais conhecida ao modelo \sig \'e o termo de Wess-Zumino-Witten (WZW) \cite{Witten2, Witten1, hull}, que, \'e um termo topol\'ogico, uma vez que n\~ao depende da m\'etrica $\gamma$. Seguindo a mesma nota\c{c}\~ao adotada at\'e aqui podemos definir este termo como:

\begin{equation} \label{eq.wz}
{\cal S}_{WZW}=\al\int d^nx\epsilon^{\mu_1\mu_2...\mu_n}\mr{b}_{i_1i_2...i_n}(\phi)\del_{\mu_1}\phi^{i_1}...\del_{\mu_n}\phi^{i_n},
\end{equation}
onde $\epsilon^{\mu_1\mu_2...\mu_n}$ \'e o tensor de Levi-Civita para o espa\c{c}o \N, $\mr{b}_{i_1i_2...i_n}$ \'e um tensor an\-tis\-si\-m\'e\-tri\-co em \M e $\al$ \'e uma constante de acoplamento.

Observemos que se $dim {\cal N}>dim{\cal M}$, teremos \'{\i}ndices repetidos em b, o que causaria uma simetria nas derivadas, anulando o produto como um todo. Ent\~ao teremos que ter: $dim {\cal N}\leq dim{\cal M}$.

As equa\c{c}\~oes de movimento podem ser achadas calculando: $\frac{\delta{\cal S}_{WZ}}{\delta\phi^l}=0$, o que nos d\'a:
\begin{equation} \label{eq.wzmov}
(n+1)\epsilon^{\mu_1...\mu_n}H_{li_1...i_n}\del_{\mu_1}\phi^{i_1}...\del_{\mu_n}\phi^{i_n}=0,
\end{equation}
onde 
\beq \label{eq.H}
H_{li_1...i_n}=\al\mr{b}_{[i_1...i_n,l]}.
\eeq
Uma vez obtidas as equa\c{c}\~oes de movimento do termo de WZW, podemos calcular as equa\c{c}\~oes de movimento da a\c{c}\~ao total ${\cal S}+{\cal S}_{WZ}$, que \'e a soma da Eq.~(\ref{eq.sigmamov}) com meia vez a Eq.~(\ref{eq.wzmov}). Este fator meio vem do termo sim\'etrico $\del_\mu\del^\mu\mr{g}_{ij}$ que n\~ao est\'a expresso na equa\c{c}\~ao acima.

Em duas dimens\~oes, podemos escrever a a\c{c}\~ao completa do modelo \sig  como sendo:
\beq
\s=\int d^2x\left(\sqrt{\gamma}\gamma^{\mu\nu}\mr{g}_{ij}+\al\epsilon^{\mu\nu}\mr{b}_{ij}\right)\del_\mu\phi^i\del_\nu\phi^j
\eeq
e suas equa\c{c}\~oes de movimento escrevem-se como:
\beq \label{eq.movsigma2}
\del_\mu(\del^\mu\phi^k)+(\Gamma^k_{ij}+\frac{3}{2}H^k_{ij})(\del_\mu\phi^i)(\del^\mu\phi^j)=0,
\eeq
onde $\Gamma^k_{ij}$ \'e o s\'{\i}mbolo de Christoffel definido anteriormente.

Observe que a Eq.(\ref{eq.wzmov}) \'e invariante pela transforma\c{c}\~ao:
$$\delta \mr{b}_{i_1...i_n}=\del_{[i_1}\lambda_{i_2...i_n]}.$$
Assim teremos: 
$$H'_{ii_1...i_n}=\del_{[i}\mr{b}_{i_1...i_n]}+\del_{[i}\del_{i_1}\lambda_{i_2...i_n]},$$ onde vemos que o segundo termo se anula, o que caracteriza uma liberdade de gauge que a a\c{c}\~ao e as equa\c{c}\~oes de movimento apresentam.

\section{Topologia.}\label{sec.top}

Fa\c{c}amos aqui uma breve discuss\~ao sobre a constante de acoplamento do termo de WZW.

Definamos uma variedade $B$ n\~ao-compacta tal que \N seja sua fronteira, ou seja, $\del B=\N$. O termo de WZW, definido em (\ref{eq.wz}) pode ser transformado, via o teorema de Stokes, em:
\begin{eqnarray}\label{eq.wzH}
\lefteqn{{\cal S}_{WZW}=\al\int_{{\cal N}=\del B} d^nx\epsilon^{\mu_1\mu_2...\mu_n}\mr{b}_{i_1i_2...i_n}(\phi)\del_{\mu_1}\phi^{i_1}...\del_{\mu_n}\phi^{i_n}=}\nonumber\\
&=&\al\int_{B} d^{n+1}x\epsilon^{\mu_1\mu_2...\mu_{n+1}}H_{i_1i_2...i_{n+1}}(\phi)\del_{\mu_1}\phi^{i_1}...\del_{\mu_{n+1}}\phi^{i_{n+1}}.
\end{eqnarray}

Observemos por\'em que os campos $\phi$ est\~ao mal definidos, uma vez que estes s\'o existem em \N e n\~ao em $B$. Devemos ent\~ao extender o mapa $\phi$ para um $\tilde{\phi}$ de forma a mapear $B$ em \M. Para que possamos fazer isto para qualquer $\phi$ e $\tilde{\phi}$, precisamos de que o n-\'esimo grupo de homotopia de \M seja trivial\footnote{O n-\'esimo grupo de homotopia de \M, ($\pi_n(\M)$, \'e definido como o conjunto dos mapeamentos, inequivalentes por qualquer tipo de deforma\c c\~ao, de uma esfera $S^n$ na variedade \M. A opera\c c\~ao deste grupo \'e a soma dos mapeamentos. Por exemplo, em  uma variedade $\RR^n$, todos os mapeamentos s\~ao homotopicamente iguais a um ponto. Se tivermos um buraco na variedade, por\'em, teremos que os mapeamentos que envolvem o buraco ser\~ao inequivalentes aos que n\~ao o envolvem e tamb\'em o ser\~ao entre s\'{\i}, pois um mapeamento que envolva duas vezes o buraco \'e inequivalente a um mesmo que envolva apenas uma vez.}
\beq
\pi_n(\M)=0.
\eeq
Isto \'e importante pois caso $\pi_n(\M)\neq 0$, ter\'{\i}amos fronteiras em \M que poderiam impedir a extens\~ao do $\phi$.

Assim, chamando $x^{n+1}=t$, teremos que $\phi=\phi(x,t)$\footnote{Em situa\c{c}\~oes onde $dH\neq 0$ para algum ponto do espa\c{c}o, \'e prefer\'{\i}vel definir o modelo de WZW atrav\'es da Eq.~(\ref{eq.wzH}) pois nestes pontos, $\mr{b}$ \'e mal definido.}. 

Fisicamente o valor do termo de WZW $(S_{WZW})$ deve ser independente da escolha de $B$, ent\~ao podemos escolher outra variedade $B'$ que tamb\'em tenha \N como sua fronteira, $\del B'=\N$, e que satisfa\c{c}a $B-B'=S^{n+1}$, ou seja, topologicamente, $B$ e $B'$ s\~ao metades de uma $(n+1)$-esfera.

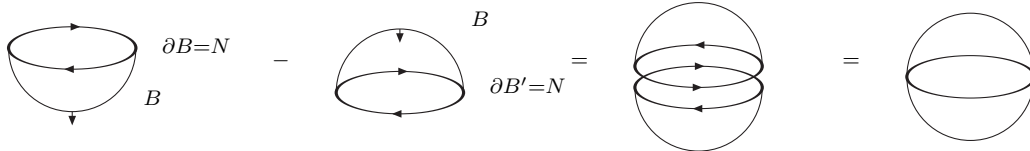
\parbox{70pt}{
\begin{picture}(100,70)(-20,-35)  
\SetColor{Black} 
\SetScale{.8} 
\Oval(10,5)(10,30)(360)
\CArc(10,5)(30,180,0)
\ArrowLine(10,15)(13,15)
\ArrowLine(10,-5)(7,-5)
\LongArrow(10,-25)(10,-30)
\Text(42,5)[lc]{$\ST \partial B=N$}
\Text(35,-15)[lc]{$\ST B$}
\end{picture}}
\hspace{1cm} $\ST -$ \hspace{1cm} 
\parbox{70pt}{
\begin{picture}(70,70)(0,-10) 
\SetColor{Black}
\SetScale{.8}
\Oval(10,15)(10,30)(360)
\CArc(10,15)(30,0,180)
\ArrowLine(10,25)(13,25)
\ArrowLine(10,5)(7,5)
\LongArrow(10,45)(10,40)
\Text(42,15)[lc]{$\ST \partial B'=N$}
\Text(35,40)[lc]{$\ST B$}
\end{picture} }  $\ST =$ \hspace{1cm}
\parbox{60pt}{
\begin{picture}(30,30)(0,0) 
\SetColor{Black}
\SetScale{.8}
\Oval(10,5)(10,30)(360)
\Oval(10,15)(10,30)(360)
\CArc(10,5)(30,180,0) 
\CArc(10,15)(30,0,180)
\ArrowLine(11,25)(10,25)
\ArrowLine(9,5)(10,5)
\ArrowLine(10,15)(11,15)
\ArrowLine(10,-5)(9,-5)
\end{picture}} $\ST =$ \hspace{1cm} 
\parbox{70pt}{
\begin{picture}(30,30)(0,0)
\SetColor{Black}\SetScale{.8}
\Oval(10,10)(10,30)(360)
\Oval(10,10)(30,30)(360) 
\end{picture}}.

Chamemos $\al\Gamma_{B}$ o segundo membro da Eq.(\ref{eq.wzH}), e $\al\Gamma_{B'}$ o mesmo para a variedade $B'$. Quanticamente, as grandezas f\'{\i}sicas surgem a partir da integrais do tipo $\int {\cal D}\phi e^{i \Gamma_S}...$, onde inserimos os campos que desejarmos na integral. Para que a f\'{\i}sica do modelo n\~ao dependa da escolha da extens\~ao de \N, teremos que ter
\beq
e^{i\al\Gamma_B}=e^{i\al\Gamma_{B'}},
\eeq
ou,
\beq
e^{i\al(\Gamma_B-\Gamma_{B'})}=e^{i\al\Gamma_S}=1,
\eeq
onde $S$ \'e a esfera $B-B'$.
Desta forma podemos concluir que,
\beq\label{eq.inteiro}
\al\Gamma_{S^{n+1}}=2\pi l,
\eeq
onde $l$ \'e um inteiro qualquer. 

Observe agora que se
\beq
\pi_{n+1}(\M)=0,
\eeq
teremos que o mapa $\tilde{\phi}$, que determina $S=\tilde{\phi}(B-B')$, pode ser extendido ainda mais, para uma variedade $(n+2)$-dimensional, $V$. Utilizando novamente o teorema de Stokes, poderemos fazer
\beq
\int_{S=\del V} H = \int_V dH = 0\mbox{, pois }dH=ddb=0.
\eeq
Por outro lado, se 
\beq
\pi_{n+1}(\M)\neq 0,
\eeq
significa que em $V$, o interior de $S$, existem ``buracos'' $(n+1)$-dimensionais, ou seja, $S$ n\~ao \'e a \'unica fronteira de $V$, teremos que
\beq
\del V=B+R_m,
\eeq
onde $R_m$ \'e a fronteira do $m$-\'esimo buraco. O teorema de Stokes fica ent\~ao alterado para 
\beq
\int_V dH=\int_S H + \int_{R_m} H = 0\mbox{, ou, }\int_S H = -\int_{R_m} H,
\eeq
 que em geral n\~ao \'e zero. 
Definamos 
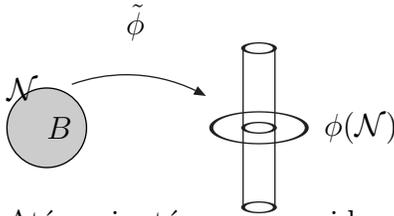
\beq
\int_S H = \Gamma_S. 
\eeq

\parbox{100pt}{
\begin{picture}(60,60)(-35,0) 
\GOval(-20,30)(15,15)(360){.8}
\LongArrowArcn(10,-10)(60,110,60)
\Oval(60,0)(2,6)(360)
\Oval(60,60)(2,6)(360)
\Line(54,0)(54,60)
\Line(66,0)(66,60)
\Oval(60,30)(2,6)(360)
\Oval(60,30)(6,18)(360)
\Text(-35,45)[lc]{$\cal N$}
\Text(-20,30)[lc]{$B$}
\Text(10,70)[lc]{$\tilde{\phi}$}
\Text(85,30)[lc]{$\phi(\cal N)$}
\end{picture}}

At\'e aqui est\'avamos considerando que $S$ pertencesse a primeira classe de equival\^encia do grupo $\pi_{n+1}(\M)$ de homotopia, ou seja, que $S$ envolvesse os buracos apenas uma vez. De maneira geral, $S$ pode pertencer a outras classes de equival\^encia, ou seja, $S$ pode envolver multiplamente os buracos. Representando por um \'{\i}ndice $q$ o n\'umero de voltas, podemos dizer que topologicamente
\beq
S_q=\underbrace{S_1+...+S_1}_{q}
\eeq
e assim teremos que
\beq
\int_{S_q}H=q\int_{S_1}H=q \Gamma_S . 
\eeq

Retornando a Eq.~(\ref{eq.inteiro}), teremos que $\al q \Gamma_S = 2\pi l$, como $l$ representa um inteiro qualquer e $q$ adv\'em do mapa $\tilde{\phi}$ que n\~ao \'e \'unico, teremos ent\~ao que ter $\al$ igual a  
\beq
\al=\frac{2\pi}{\Gamma_S}n,
\eeq
com n inteiro,
ent\~ao a Eq.~(\ref{eq.wzH}) fica corretamente escrita na forma: 
\beq
S_{WZW}=\frac{2\pi}{\Gamma_S}n\Gamma_{B}.
\eeq


%

\section{Modelo \sig em Variedades de Grupos}\label{sec.modgrupo}

Apresentemos agora uma outra defini\c c\~ao do modelo sigma encontrada freq\"uentemente, \cite{Witten2, Witten1}, que est\'a relacionada diretamente com variedades de grupos.

Seja um grupo de Lie {\got G}, e sejam $g$ seus elementos: $g\in \mbox{\got G}$. Definamos um mapa $g=g(x)$ que relaciona \N com a variedade de grupo de {\got G}. Este mapa pode ser implementado via os vetores $\phi$ da \'algebra {\got g} atrav\'es do mapa $\phi=\phi(x)$. Ent\~ao, escrevamos
\beq
g=e^\phi=e^{\mb{T_i}\phi^i},
\eeq
onde $\mb{T_i}$ s\~ao os geradores da \'algebra. Dessa forma, a a\c{c}\~ao do modelo \sig  n\~ao-linear \'e definida como:
\beq\label{eq.sigmagrupo}
\s=\int d^nx\sqrt{\gamma} \gamma^{\mu\nu}\tr \left(g^{-1}\del_\mu gg^{-1}\del_\nu g\right),
\eeq
cujas equa\c{c}\~oes de movimento s\~ao:
\beq
\gamma^{\mu\nu}g^{-1}\del_\mu\left(\del_\nu g g^{-1}\right)=0.
\eeq


Introduzamos o termo de WZW. Como dito na se\c{c}\~ao \ref{sec.invariancias}, este \'e um termo topol\'ogico e para defini-lo, fa\c{c}amos semelhantemente como na se\c{c}\~ao \ref{sec.top}. 

Sabemos que $g(x)$ determina uma imers\~ao de \N na variedade de grupo {\got G}. Fa\c{c}amos uma extens\~ao de \N para uma variedade $B$, tal que $\del B=\N$. Isto implicar\'a em uma extens\~ao $\tilde{g}(B)$ se tivermos que
\beq
\pi_n(\mbox{\got G})=0.
\eeq
Assim, por exemplo, se $\N=S^n$, $B$ ser\'a uma meia esfera $S^{n+1}$. Escolhendo $t$ como a $(n+1)$-\'esima coordenada, podemos definir $\tilde{g}$ a satisfazer por exemplo
\beq
\tilde{g}=\tilde{g}(x,t),\quad \tilde{g}(0,x)=1,\quad \tilde{g}(1,x)=g(x),
\eeq
ou explicitamente, $\tilde{g}=e^{\phi^iT_it}$.

Tendo isto, definamos uma (n+1)-forma:   
\beq\label{eq.Tn}
{\cal T}_{\mu_1...\mu_{n+1}}=\tr\left[(\tilde{g}^{-1}\del_{\mu_1}\tilde{g})...(\tilde{g}^{-1}\del_{\mu_{n+1}}\tilde{g})\right],
\eeq
definida em $\tilde{g}(S^{n+1})$. O termo de WZW \'e tido como o ``fluxo'' desta forma atrav\'es de $B$, \cite{Witten2}.
\beq\label{eq.wzwgrupo}
\s_{WZW}=\int_{B} d^{n+1}x\epsilon^{\mu_1...\mu_{n+1}}{\cal T}_{\mu_1...\mu_{n+1}},
\eeq
Pelos argumentos dados na se\c{c}\~ao \ref{sec.top}, se $\pi_{n+1}(\mbox{\got G})\neq 0$, a constante de acoplamento de $S_{WZW}$ dever\'a ser um m\'ultiplo inteiro de $\frac{2\pi}{m}$, onde $m$ \'e a integral (\ref{eq.wzwgrupo}) feita sobre a esfera $S^{n+1}$.

Assim, em $n$ dimens\~oes o modelo \sig  se escreve como:
\beq\label{eq.modsigmagrupo}
\s=\int d^nx\tr \left(g^{-1}\del_\mu g g^{-1}\del^\mu g\right)+\al\int d^nxdt \epsilon^{\mu_1...\mu_{n+1}}{\cal T}_{\mu_1...\mu_{n+1}}.
\eeq
Em duas dimens\~oes de espa\c{c}o-tempo, teremos:
\beq\label{eq.sigmagrupo2}
\s=\int d^2x\tr \left(g^{-1}\del_\mu g g^{-1}\del^\mu g\right)+\al\int d^2xdt \epsilon^{\mu\nu\rho}\tr \left(g^{-1}\del_\mu g g^{-1}\del_\nu g g^{-1}\del_\rho g\right).
\eeq

\subsection{Equival\^encia entre os modelos.}

Consideremos que a variedade alvo, \M, possua isometrias que definem um grupo {\got G}. Observe que este grupo tera $\frac{n(n+1)}{2}$ par\^ametros, uma vez que este \'e o n\'umero de isometrias do espa\c co $\RR^n$ euclidiano que jutamente com os espa\c cos de De Sitter e Anti-De Sitter s\~ao os mais sim\'etricos poss\'{\i}veis.
Seja o grupo {\got G}, com elementos $g$, e par\^ametros $\phi^a$. A m\'etrica desta variedade \'e definida como \cite{Spindel}:
\beq \label{eq.metricagrupo}
\mr{g}_{ab}=\tr((g^{-1})_{,a}g_{,b}),
\eeq
onde $\DS g_{,a}=\frac{\DS \del g}{\DS \del \phi^a}$.

Esta defini\c{c}\~ao claramente associa a Eq.~(\ref{eq.sigmagrupo}) com a a\c{c}\~ao (\ref{eq.acao}), pois:
\beq
\gamma^{\mu\nu}\mr{g}_{ab}\del_\mu\phi^a\del_\nu\phi^b=\gamma^{\mu\nu}\tr(g^{-1}_{,a}g_{,b})\del_\mu\phi^a\del_\nu\phi^b=\tr(\del_\mu g^{-1}\del_\nu g).
\eeq

Quanto ao termo de WZW, podemos associar a defini\c{c}\~ao para variedades de grupo, Eq.~(\ref{eq.wzwgrupo}), e a defini\c{c}\~ao geral, Eq.~(\ref{eq.wzH}) se fizermos a identifica\c{c}\~ao:
\beq
H_{a_1...a_{n+1}}=\tr \left[(g^{-1}g,_{a_1})...(g^{-1}g,_{a_{n+1}})\right],
\eeq
pois assim,
\beq
H_{a_1...a_{n+1}}\del_{\mu_1}\phi^{i_1}...\del_{\mu_{n+1}}=\tr \left[(g^{-1}g,_{\mu_1})...(g^{-1}g,_{\mu_{n+1}})\right]={\cal T}.
\eeq

Dessa forma, teremos:
\begin{eqnarray}
\s_{WZW}&=&q\int_Bd^{n+1}x\epsilon^{\mu_1...\mu{n+1}}H_{a_1...a_{n+1}}\del_{\mu_1}\phi^{a_1}...\del_{\mu_{n+1}}\phi^{a_{n+1}}=\nonumber\\
&=&q\int_Bd^{n+1}x\epsilon^{\mu_1...\mu{n+1}}{\cal T}_{\mu_1...\mu_{n+1}}.
\end{eqnarray}

Uma associa\c{c}\~ao mais clara e geom\'etrica entre as defini\c{c}\~oes do termo de WZW, que sera feita a seguir, \'e encontrada quando analisamos o caso bi-dimensional (n=2), onde teremos uma rela\c{c}\~ao entre b, $g$, e a tor\c{c}\~ao do espa\c{c}o-tempo.

\section{Tetrada e Tor\c{c}\~ao}\label{sec.tetrada}

Para que possamos mostrar a rela\c{c}\~ao entre as duas formula\c{c}\~oes do modelo $\sig$, \'e preciso discutir alguns t\'opicos de geometria diferencial. Precisamos inicialmente definir uma m\'etrica sobre {\got G} de forma que possamos calcular grandezas geom\'etricas como curvatura.  

Nesta se\c c\~ao usaremos a seguinte nota\c c\~ao:
\begin{itemize}
\item S\'{\i}mbolos em negrito ou com \'{\i}ndices em negrito: tensores. (p.ex.: $\mb{T_i}$)
\item S\'{\i}mbolos com \'{\i}ndices em it\'alico: componentes de tensores. (p.ex.: $V_a$)
\item M\'etrica: $\GG{A}{B}=\mr{g}_{ab}A^aB^b$.
\item Derivada de Lie entre dois vetores: \lie{A}{B}.
\item Derivada Covariante: $(\DD{A}{B})^b=A^a\nabla_aB^b=A^aB^b{}_{;a}$, onde o s\'{\i}mbolo $;a$ indica a derivada covariante com respeito a $a$.  
\item A Conex\~ao em uma base de vetores $\mb{E_a}$ \'e definida como: $\DD{{E_b}}{E_c}=\Gamma^a_{bc}E_a$.
\item \'E importante lembrar que em geometria diferencial os vetores comportam-se como diferenciais.
\item \'Indices de a-h s\~ao do espa\c co-tempo e \'{\i}ndices i-z representam espa\c co tangente.
\end{itemize}

Observemos que, na \'algebra {\got g}, os vetores da base, $\mb{T_i}$, satisfazem uma rela\c{c}\~ao de comuta\c{c}\~ao:
\beq
[\mb{T_i},\mb{T_j}]=C_{ij}^k\mb{T_k},
\eeq
onde $C_{ij}^k$ s\~ao as constantes de estrutura do grupo {\got G}. Como se sabe da geometria diferencial, uma base de vetores que n\~ao comutam n\~ao pode definir uma base coordenada, uma vez que vetores coordenados comutam. Ademais, como a \'algebra \'e na realidade o espa\c{c}o vetorial tangente \`a variedade, sua m\'etrica \'e constante. Define-se a m\'etrica de uma \'algebra como, \cite{sattinger}, 
\beq
\eta_{ij}=\tr(\mb{T_i}\mb{T_j}).
\eeq
Definamos a base tetrada $T_a{}^i$ que encarrega-se de levar componentes do espa\c co curvo para componentes da base ortonormal tangente, a \'algebra. Assim,
\beq\label{eq.tetrada1}
g^{-1}\del_a g=\Lambda_a{}^i\mb{T_i}.
\eeq
Dessa maneira:
\beq
g^{-1}\del_\mu g=g^{-1}\del_a g\del_\mu\phi^a=\mb{T_a}\del_\mu \phi^a,
\eeq
 onde $\mb{T_a}=\Lambda_a{}^i\mb{T_i}$, que \'e um vetor de uma base coordenada, n\~ao necessariamente ortonormal. Assim, a lagrangeana do modelo \sig principal se escrever\'a como:
\begin{eqnarray}\label{eq.sigmaprincipal}
\tr(g^{-1}\del_\mu gg^{-1}\del^\mu g)&=&\tr(\mb{T_aT_b})\del_\mu\phi^a\del^\mu\phi^b=\nonumber\\
&=&\mr{g}_{ab}\del_\mu\phi^a\del^\mu\phi^b,
\end{eqnarray}
onde podemos ver que $\mr{g}_{ab}=\Lambda_a{}^i\Lambda_b{}^j\eta_{ij}$. Esta \'e a lagrangeana da a\c{c}\~ao (\ref{eq.sigmagrupo}).

Em resumo, o que temos \'e o seguinte. O espa\c{c}o tangente da variedade do grupo {\got G} \'e a \'algebra {\got g}. Tomando seus geradores como os vetores da base, teremos que estes n\~ao comutam. Isto \'e equivalente a dizer que a derivada de Lie entre eles \'e n\~ao nula: $\lie{T_i}{T_j}\neq 0$, e como a m\'etrica deste espa\c{c}o \'e constante, podemos interpretar os vetores da base como formadores de uma base tetrada. Existe por\'em uma outra base vetorial cuja m\'etrica \'e definida pela Eq.~(\ref{eq.metricagrupo}). Para fazer a transi\c c\~ao entre estas bases, definimos a tetrada $T_a{}^i$, Eq~(\ref{eq.tetrada1}).


Para discutirmos a interpreta\c{c}\~ao do termo de WZW, precisamos discutir a quest\~ao da tor\c{c}\~ao na defini\c{c}\~ao das conex\~oes.
O Tensor de Tor\c{c}\~ao \'e definido atrav\'es de\footnote{A nota\c{c}\~ao aqui segue semelhante \`a \cite{hawking}}:
\beq
\Tor{X}{Y}=\DD{X}{Y}-\DD{Y}{X}-\lie{X}{Y}=\DD{X}{Y}-\DD{Y}{X}-[\mb{X},\mb{Y}],
\eeq
que \'e antissim\'etrico em $\mb{X}$ e $\mb{Y}$. Podemos obter suas componentes observando que em termos dos vetores da base $\mb{E_i}$ pode ser escrito na forma:
\beq
\Tor{X}{Y}=T^i_{jk}X^jY^k\mb{E_i}
\eeq
e teremos a antissimetria representada por: $T^i{}_{jk}=T^i{}_{[kj]}$.

Usualmente, a defini\c{c}\~ao da derivada covariante, logo da conex\~ao, exige que \Tor{X}{Y}=0 para qualquer vetor. Por\'em esta condi\c{c}\~ao ser\'a relaxada pois existe uma estreita rela\c c\~ao entre a tor\c{c}\~ao e o termos de WZW em duas dimens\~oes.
A outra condi\c{c}\~ao usual, que ser\'a mantida \'e a const\^ancia covariante da m\'etrica, a saber:
\beq
\DD{X}{\mr{g}}=0.
\eeq

Como estamos procurando as componentes da conex\~ao, calculemos, $\GG{Z}{\DD{X}{Y}}$, que ap\'os f\'acil manipula\c{c}\~ao fica:
\begin{eqnarray}
\GG{Z}{\DD{X}{Y}}&=&\frac{1}{2}\left\{ \mb{X}[\GG{Z}{Y}]+\mb{Y}[\GG{X}{Z}]-\mb{Z}[\GG{X}{Y}]+\right.\nonumber\\
&&{}+\GG{\DD{Z}{X}-\DD{X}{Z}}{Y})+\GG{\DD{Z}{Y}-\DD{Y}{Z}}{X}+\nonumber\\
&&{}+\left.\GG{\DD{X}{Y}-\DD{Y}{X}}{Z}\right\}.
\end{eqnarray}
Utilizando a tor\c{c}\~ao, teremos:
\begin{eqnarray}
\GG{Z}{\DD{X}{Y}}&=&\frac{1}{2}\left\{\mb{X}[\GG{Z}{Y}]+\mb{Y}[\GG{X}{Z}]-\mb{Z}[\GG{X}{Y})] +\right.\nonumber\\
&&\GG{[Z,X]+\Tor{Z}{X}}{Y})+\GG{[Z,Y]+\Tor{Z}{Y}}{X})+\nonumber\\
&&\left.\GG{[X,Y]+\Tor{X}{Y}}{Z})\right\}.
\end{eqnarray}
Para recobrar a rela\c{c}\~ao usual sem tor\c{c}\~ao, como em \cite[p. 40]{hawking}, basta fazer \Tor{X}{Y}=0.

Substituindo os vetores $\mb{X}$,$\mb{Y}$, $\mb{Z}$, por vetores de uma base, $\mb{E_n}$, teremos:
\begin{eqnarray}
\lefteqn{\GG{E_a}{\DD{E_b}{E_c}}=\frac{1}{2}\left\{\mb{E_b} \GG{E_a}{E_c}+ \mb{E_c} \GG{E_b}{E_a}- \mb{E_a} \GG{E_b}{E_c}+\right.}\nonumber\\
&&\left.{}+C_{cab}+C_{bac}+C_{abc}+T_{cab}+T_{bac}+T_{abc}\right\}=\Gamma_{abc},
\end{eqnarray}
onde aqui definimos, 
\begin{eqnarray}
\GG{E_c}{[E_a,E_b]}=C^d{}_{ab}\GG{E_c}{E_d}=\mr{g}_{cd}C^d{}_{ab}=C_{cba}\\
\GG{\Tor{E_a}{E_b}}{E_c}=T^d{}_{ab}\GG{E_d}{E_c}=\mr{g}_{dc}T^d{}_{ab}=T_{cab}.
\end{eqnarray}

Em uma base tetrada, onde a m\'etrica \'e ortogonal e constante, $\mb{E_i}\GG{E_j}{E_k}=0$, ent\~ao teremos:
\beq
\Gamma_{ijk}=\frac{1}{2}\left\{\eta_{kl}C^l{}_{ij}+\eta_{jl}C^l{}_{ik}+\eta_{il}C^l{}_{jk}+T_{kij}+T_{jik}+T_{ijk}\right\}.
\eeq
Em uma base coordenada:
\beq
\Gamma_{abc}=\frac{1}{2}\left\{\del_b\mr{g}_{ac}+\del_c\mr{g}_{ab}-\del_a\mr{g}_{bc}+T_{abc}+T_{bac}+T_{cab}\right\}.
\eeq

Tentemos agora encontrar as componentes da tor\c{c}\~ao. Calculemos a derivada covariante da base tetrada:
\beq
\DD{E_b}{T_i}=\DD{\del_b}{(\mbox{$\Lambda_i{}^a$}E_a)}
\eeq
e tamb\'em temos que:
\beq
\DD{E_b}{T_i}=\DD{{\mbox{$\Lambda_b{}^j$}}T_j}{T_i}=\Lambda_b{}^j\Lambda_k{}^c\Gamma^k_{ji}\mb{E_c}.
\eeq
Ent\~ao:
\beq
\DD{E_b}{(\mbox{$\Lambda_i{}^a$}E_a)}=\Lambda_b{}^j\Lambda_k{}^c\Gamma^k_{ji}\mb{E_c}.
\eeq

\'E interessante para variedades de grupo estudar apenas o caso da tor\c{c}\~ao completamente antissim\'etrica. Ent\~ao, admitindo que 
\beq
T_{ijk}=T_{[ijk]},
\eeq
poderemos escrever as componentes da derivada da base tetrada como sendo:
\beq
\Lambda_i{}^a{}_{;b}=\frac{1}{2}\left\{\Lambda_b{}^j\Lambda_k{}^aC^k{}_{ji}+T_{ijk}\eta^{mj}\Lambda_b{}^k\Lambda_m{}^a\right\}.
\eeq
Para chegarmos em nosso objetivo, que \'e relacionar a tor\c{c}\~ao com o termo de WZW, exijamos que a tetrada seja uma constante covariante, ou seja,
\beq\label{eq.integrtetra}
\Lambda_i{}^a{}_{;b}=0,
\eeq
o que implicar\'a em
\beq\label{eq.TC}
T^a{}_{bc}=\Lambda_b{}^i\Lambda_c{}^j\Lambda_k{}^aC^k_{ij}.
\eeq

Observe agora que
\begin{eqnarray}
\tr(g^{-1}g_{,[a}g^{-1}g_{,b}g^{-1}g_{,c]})&=&\tr(\mb{T_{[a}T_{b}T_{c]}})=\Lambda_a{}^i\Lambda_b{}^j\Lambda_c{}^k\tr(\mb{T_{[i}T_{i}T_{k]}})=\nonumber\\
&=&\frac{1}{6}\Lambda_a{}^i\Lambda_b{}^j\Lambda_c{}^k\tr(C^l_{ij}\mb{T_lT_k}+C^l_{jk}\mb{T_lT_i}+C^l_{ki}\mb{T_lT_j})=\nonumber\\
&=&\frac{1}{6}\Lambda_a{}^i\Lambda_b{}^j\Lambda_c{}^k(C_{kij}+C_{ijk}+C_{jki})=\nonumber\\
&=&\frac{1}{2}\Lambda_a{}^i\Lambda_b{}^j\Lambda_c{}^kC_{ijk}.
\end{eqnarray}
Assim, 
\beq\label{eq.Tg}
T_{abc}=2 \tr(g^{-1}g_{,[a}g^{-1}g_{,b}g^{-1}g_{,c]}).
\eeq

Ou seja, em uma variedade de grupo, se impusermos a condi\c{c}\~ao de ``integrabilidade'' da tetrada, Eq.~(\ref{eq.integrtetra}), e exigirmos a antissimetria da tor\c{c}\~ao, teremos a mesma definida pelas Eq.~(\ref{eq.TC}) ou Eq.~(\ref{eq.Tg}).


J\'a sabemos, das equa\c{c}\~oes de movimento, Eq.~(\ref{eq.movsigma2}), que $3 H^k_{ij}$ cumpre o papel da tor\c{c}\~ao. Em uma variedade de grupo, a tor\c{c}\~ao \'e definida atrav\'es da equa\c{c}\~ao acima. Para que possamos interpretar o termo de WZW definido em Eq.~(\ref{eq.wzH}) como sendo a tor\c{c}\~ao do espa\c{c}o-tempo, teremos que definir
\beq\label{eq.WZWH}
H_{abc}=\frac{2}{3}{\cal T}_{abc}.
\eeq

\subsection{Invari\^ancia Conforme}

Um dos aspectos mais importantes do modelo sigma \'e a an\'alise de seus pontos cr\'{\i}ticos, em particular, sabe-se que modelos sigma em duas dimens\~oes podem ser ajustados a fim de que estes sejam finitos, ou seja, com fun\c c\~ao beta nula. Uma analise deste assunto pode ser vista em \cite{Witten1}. Recapitulando esta quest\~ao, temos que o modelo sigma da forma geral
\beq\label{eq.acaoinva}
S=\frac{1}{4\lambda^2}\int d^2x \tr \left(\del_\mu g\del^\mu g^{-1}\right)+\frac{n}{24\pi}\int d^3y\epsilon^{\mu\nu\rho}\tr\left(g^{-1}\del_\mu g g^{-1}\del_\nu g g^{-1}\del_\rho g\right), 
\eeq
apresenta invari\^ancia conforme no ponto $\lambda^2=\frac{4\pi}{n}$. Neste ponto, podemos escrever as equa\c c\~oes de movimento de maneira mais simples. No cone de luz teremos
\beq\label{eq.estrela}
\del_-\left(g^{-1}\del_+g\right)=0\mbox{, para }\lambda^2>0\mbox{ e }\del_+\left(g^{-1}\del_-g\right)=0\mbox{, para }\lambda^2<0.
\eeq

Para chegar a estas equa\c c\~oes, devemos calcular a varia\c c\~ao da a\c c\~ao com respeito aos campos $g$. O termo principal \'e relativamente simples e lembrando que $\delta g^{-1}=-g^{-1}\delta g g^{-1}$ teremos
\beq
\delta \left(\del_\mu g\del^\mu g^{-1}\right)=\left( \delta g \del_\mu(g^{-1}\del^\mu g)g^{-1}\right).
\eeq

O termo de WZW \'e mais interessante pois \'e topol\'ogico e vive num espa\c co de tr\^es dimens\~oes. Observe que ap\'os integra\c c\~ao por partes, a sua varia\c c\~ao assume a forma
\begin{eqnarray}
\delta \Gamma&=&3\int d^3y\epsilon^{\mu\nu\rho}\tr \del_\mu \left(g^{-1}\delta g g^{-1}\del_\nu g g^{-1} \del_\rho g\right)\nonumber\\
&=&3\int d^2x\epsilon^{\mu\nu}\tr \left(g^{-1}\delta gg^{-1}\del_\mu gg^{-1}\del_\nu g\right)=\nonumber\\    
&=&-3\int d^2x\epsilon^{\mu\nu}\tr \left[g^{-1}\delta g\del_\mu \left(g^{-1}\del_\nu g\right)\right].
\end{eqnarray}

Desta maneira, podemos escrever as equa\c c\~oes de movimento da a\c c\~ao (\ref{eq.acaoinva}) como
\beq
\frac{1}{2\lambda^2}\del_\mu\left(g^{-1}\del^\mu g\right)g^{-1}-\frac{n \epsilon^{\mu\nu}}{8\pi}\del_\mu\left(g^{-1}\del^\mu g\right)=0.
\eeq
Como no cone de luz $\mr{g}=\left(\begin{array}{rr}0&1\\1&0\end{array}\right)$ e $\epsilon=\left(\begin{array}{rr}0&1\\-1&0\end{array}\right)$, teremos que no ponto cr\'{\i}tico, as equa\c c\~oes de movimento ser\~ao exatamente as descritas em (\ref{eq.estrela}).

Desejo aqui ressaltar dois detalhes. Primeiramente, no ponto cr\'{\i}tico, a raz\~ao entre os coeficientes do termo principal e do termo de WZW \'e $\frac{2}{3}$, ou seja, esta \'e a raz\~ao para que tenhamos invari\^ancia conforme, considerando que esta n\~ao \'e alterada por um termo global na a\c c\~ao.
Observemos agora que quando temos a tor\c c\~ao, da Eq.~(\ref{eq.movsigma2}), igual a tor\c c\~ao da variedade de grupo, Eq.~(\ref{eq.Tg}), as equa\c c\~oes de movimento tornam-se 
\beq \label{eq.duasestrelas}
\del_+\del_-\phi^a+(\Gamma^a{}_{bc}+{\cal T}^a_{bc})\del_+\phi^b\del_-\phi^c=0.
\eeq
Para que possamos relacionar estas equa\c c\~oes com a Eq.~(\ref{eq.estrela}), observemos que se $g=g_0e^{\phi^i\mb{T_i}}$, onde $\phi^i$ \'e infinitesimal e $g$ est\'a pr\'oximo de $g_0$, teremos
\beq\label{eq.tresestrelas}
\del_-\left(g^{-1}\del_+g\right)=\left(\del_-\del_+\phi^k+C^k_{ij}\del_+\phi^i\del_-\phi^j\right)\tilde{g}^{-1}\mb{T_k}\tilde{g}=0.
\eeq
Observe agora que se escrevermos a Eq~(\ref{eq.duasestrelas}) utilisando a base tetrada (ortonormal da \'algebra) teremos que
\beq
\Gamma_{ijk}=\meio C_{ijk},
\eeq
como ${\cal T}_{ijk}=\meio C_{ijk}$, a Eq.~(\ref{eq.duasestrelas}) se escreve como
\beq
\del_+\del_-\phi^k+C^k_{ij}\del_+\phi^i\del_-\phi^j=0,
\eeq
que \'e a mesma equa\c c\~ao de (\ref{eq.tresestrelas}), logo para termos invari\^ancia conforme, $H=\frac{2}{3}{\cal T}$.

\section{A\c{c}\~ao Efetiva.}

A teoria de Cordas Bos\^onicas descreve uma folha mundo sobre o espa\c{c}o-tempo. Observando sua a\c{c}\~ao, v\^e-se que n\~ao existe nenhuma restri\c{c}\~ao sobre o espa\c{c}o-tempo f\'{\i}sico, na forma de que n\~ao existe nenhuma restri\c{c}\~ao sobre a m\'etrica $\mr{g_{ij}}$. Isto ocorrendo no n\'{\i}vel cl\'assico. 
Como sabemos de v\'arias teorias f\'{\i}sicas, a quantiza\c{c}\~ao via de regra nos traz ``surpresas'' com rela\c{c}\~ao \`as simetrias e leis de conserva\c{c}\~ao, muitas vezes violando-as, dando origem \`as chamadas anomalias qu\^anticas.
Para buscarmos a influ\^encia qu\^antica sobre a m\'etrica, devemos ver que ela se comporta como uma ``constante de acoplamento'' da a\c{c}\~ao bi-dimensional, e desta forma seu comportamento qu\^antico ser\'a dado pela fun\c{c}\~ao $\be$, que define a varia\c{c}\~ao da constante de acoplamento com o cut-off\footnote{Nesta se\c{c}\~ao usaremos os \'{\i}ndices gregos para o espa\c{c}o-tempo f\'{\i}sico (o espa\c{c}o-alvo) e latinas para a folha-mundo.}:
\beq
\be_{ab(x)}(E)=\frac{\del \mr{g}_{ab}(x)}{\del \ln E}.
\eeq
     


A a\c c\~ao de Polyakov da Corda Bos\^onica sobre uma variedade \'e \footnote{\cite{pol}, \cite{polchinski}, cap. 1.2, \cite{green}, cap. 3.4,\cite{hull}, cap. 15.}: 

\beq \label{polyakov}
S_P=\frac{1}{4\pi\alpha^\prime}\int d^2\sigma \sqrt{\mr{g}}\mr{g}^{ab}G_{\mu\nu}(X)\del_aX^\mu\del_bX^\nu.
\eeq
Onde nesta a\c c\~ao, $G_{\mu\nu}$ \'e a m\'etrica do espa\c co-tempo, $g_{ab}$ \'e a m\'etrica da folha-mundo da corda, $X^\mu$ \'e o vetor posi\c c\~ao no espa\c co-tempo e $\alpha^\prime$ \'e a ``inclina\c c\~ao de Regge''. Aqui estamos considerando que $\Phi=B_{\mu\nu}=V=0$, ou seja, estamos desprezando o d\'{\i}laton, o tensor antissim\'etrico e o termo potencial.


Fa\c camos uma expans\~ao na m\'etrica, da forma:

\beq G_{\mu\nu}(X)=\eta_{\mu\nu}+h_{\mu\nu}(X), \eeq 
onde $h_{\mu\nu}$ \'e pequeno, poderemos escrever o funcional gerador como:

\beq
Z_\mr{g}=\int {\cal D}\!X{\cal D}\!h\,e^{-S_{P\eta}}\left(1-\frac{1}{4\pi\alpha^\prime}\int d^2\sigma \sqrt{\mr{g}}\mr{g}^{ab}h_{\mu\nu}(X)\del_aX^\mu\del_bX^\nu+O(\alpha^{\prime 2})\right),
\eeq
onde o \'{\i}ndice $\eta$ na a\c c\~ao indica que estamos usando a m\'etrica de Minkowsky em vez da m\'etrica geral $G$. O segundo termo que aparece entre par\^enteses, \'e o operador de v\'ertice do gr\'aviton da corda fechada\footnote{Podemos fazer ${h_{\mu\nu}}=-2\pi\alpha^\prime e^{ik.X}D_{\mu\nu}$, $h_{\mu\nu}$ assume a forma de um gr\'aviton.}, $V_{graviton}$ que representa a inser\c c\~ao de um gr\'aviton\footnote{cap:3.6-\cite{polchinski}}. Considerando os termos de ordem superior em $\alpha$, teremos a exponencia\c c\~ao do operador de v\'ertice: $e^{V_{graviton}}$, que define um ``estado coerente'' de gr\'avitons. Desta maneira, \'e natural pensar que ao se utilizar a a\c c\~ao definida em~(\ref{polyakov}), teremos a descri\c c\~ao completa da intera\c c\~ao gravitacional. Ou de outra maneira, a descri\c c\~ao da corda sobre um espa\c co-tempo curvo.  

Para quantizar a a\c c\~ao de Polyakov, definamos o {\bf funcional gerador}\footnote{Como em \cite{peskin}, cap:9}:

\beq Z_J=\int {\cal D}\!Xe^{-S_P-X.J}, \eeq onde $X.J=\int d^2\sigma X^\mu J_\mu$.
Definamos o {\bf funcional energia}:

\beq
e^{-W}=Z_J\rightarrow W=-\ln Z_J, \eeq 
onde $J$ \'e uma fonte acoplada ao campo $X^\mu$, que ser\'a levada a zero no c\'alculo dos valores esperados.
Atrav\'es deste funcional, podemos definir o campo m\'edio:

\beq
\bar{X^i}(\sigma)=\frac{\delta W}{\delta J_i},\eeq pois $\frac{\delta W}{\delta J_i}=\frac{-1}{Z_J}\frac{\delta Z_J}{\delta J_i}=\frac{-1}{Z_J}\int{\cal D}\!Xe^{-(S+X.J)}(-X^i) = \left<X^i\right>$.

Definamos ent\~ao a {\bf a\c c\~ao efetiva}:
\beq \Gamma(\bar{X})=W(J)-\bar{X}.J(\bar{X}). \eeq

Este \'ultimo funcional, escrito em termos da {\bf flutua\c c\~ao qu\^antica}: $\Pi^i=X^i-\bar{X^i}$, assume a forma:

\beq
e^{-\Gamma(\bar{X})}=\int {\cal D}\Pi\,e^{-S[\bar{X}+\Pi]+\Pi.J}.
\eeq

\'E importante lembrar que a a\c c\~ao efetiva nos fornece os diagramas 1PI, ou seja
\beq
e^{-\Gamma}\int {\cal D}\Pi f(\Pi) e^{-S[\bar{X}+\Pi]+\Pi.J}=\left<f(\Pi)\right>,
\eeq
onde o valor esperado \'e calculado apenas com os diagramas irredutiveis de uma part\'{\i}cula. 

Expandindo a a\c c\~ao efetiva em s\'erie de Taylor e guardando apenas a primeira ordem em $\Pi$, poderemos escrever a a\c c\~ao como:

\beq
\Gamma[\Pi]=S[\bar{X}]+\frac{1}{4\pi\alpha^\prime}\int d^2\sigma \left[\frac{1}{2}G_{\mu\nu}\del_a\Pi^\mu\del^a\Pi^\nu+A^a_{\mu\nu}\Pi^\mu\del_a\Pi^\nu+\frac{1}{2}B_{\mu\nu}\Pi^\mu\Pi^\nu\right], \eeq onde,
$$\left\{ 
\begin{array}{rcl} 
A^a_{\mu\nu}&=&G_{\mu\rho,\nu}\del^a\bar{X}^\rho \\
B_{\mu\nu}&=&\frac{1}{2}G_{\rho\sigma,\mu\nu}\del^a\bar{X}^\sigma\del_a\bar{X}^\rho \end{array}\right.$$

Vale observar que estes campos definidos acima n\~ao s\~ao, na realidade, campos propagantes. Podem ser pensados como campos cl\'assicos ``de fundo'', que agem como fontes do campo qu\^antico.

Inserindo um regulador infravermelho, $\int d^2\!\sigma \frac{m^2}{2}\delta_{ij}\Pi^i\Pi^j$, podemos obter as regras de Feynman. Diagramaticamente:\\

\parbox{50pt}{\begin{picture}(10,40)(0,-20)
\ArrowLine(0,0)(40,0)
\Text(-3,10)[tl]{$\mu$}
\Text(43,10)[tr]{$\nu$}
\end{picture}}
\hspace{1cm} $=$ \hspace{1 cm} $\frac{\displaystyle -G_{\mu\nu}}{\displaystyle p^2+m^2}.$\hfill
\parbox{2cm}{\beq \eeq}

\vspace{12pt}

\parbox{50pt}{
\begin{picture}(50,40)(0,-20)
\Line(0,-1)(30,-1)
\Line(0,1)(30,1)
\Line(30,0)(50,20)
\Line(30,0)(50,-20)
\Vertex(30,0){2}
\Text(3,3)[bl]{$A$}
\Text(3,-3)[tl]{$a$}
\Text(50,20)[tl]{$\nu$}
\Text(50,-20)[tl]{$\mu$}
\Text(40,10)[tl]{$P_a$}
\end{picture}}
\hspace{1cm} $=$
\hspace{1cm} $A^a_{\mu\nu}P_a.$\hfill\parbox{2cm}{\beq \eeq}

\vspace{12pt}

\parbox{50pt}{
\begin{picture}(50,40)(0,-20)
\Line(0,-1)(30,-1)
\Line(0,1)(30,1)
\Line(30,0)(50,20)
\Line(30,0)(50,-20)
\Vertex(30,0){2}
\Text(3,3)[bl]{B}
\Text(50,20)[tl]{$\nu$}
\Text(50,-20)[tl]{$\mu$}
\end{picture}}\hspace{1cm} $=$\hspace{1cm} $B_{\mu\nu}.$\hfill\parbox{2cm}{\beq
\eeq}
\\
\\

Nesta teoria a {\em constante de acoplamento} \'e a m\'etrica $G_{\mu\nu}$. Para sabermos sua din\^amica, ou seja o grupo de renormaliza\c c\~ao, precisamos calcular os contra-termos e a fun\c c\~ao beta.

\parbox{2.5cm}{\bolA}\hspace{1cm} $\DS=$\hspace{1cm}$\DS-\int A^a_{\mu\nu}P_a \frac{G_{\mu\nu}}{p^2+m^2}\frac{d^2p}{(2\pi)^2}$\hfill\parbox{2cm}{\beq \eeq}
\\

\parbox{2.5cm}{\bolB}\hspace{1cm} $\DS=$\hspace{1cm}$\DS-\int B_{\mu\nu}\frac{G^{\mu\nu}}{p^2+m^2}\frac{d^2p}{(2\pi)^2}$\hfill\parbox{2cm}{\beq \eeq}
\\

\parbox{2.5cm}{\diagAA }\hspace{1cm} $\DS=$\hspace{1cm}$\DS\int A^a_{\mu\nu}P_a\frac{G^{\mu\rho}}{p^2+m^2}A^b_{\rho\sigma}(p-k)_b\times$\\
\hspace*{6cm} $\DS\times\frac{G^{\sigma\mu}}{(p-k)^2+m^2}\frac{d^2p}{(2\pi)^2}$\hfill\parbox{2cm}{\beq \eeq}

Nesta aproxima\c c\~ao, temos apenas corre\c c\~oes qu\^anticas para o v\'acuo.
Observe que o primeiro diagrama \'e nulo e nesta teoria temos dois diagramas logaritmicamente divergentes.

Calculando as integrais, utilizando os par\^ametros de Feynman e Regulariza\c c\~ao dimensional (que nos fornecer\'a o par\^ametro de massa $\mu^\epsilon$), podemos concluir que a parte divergente destes dois diagramas escreve-se como:
\\

\parbox{2.5cm}{\bolB}\hspace{-12pt}+\hspace{12pt} \parbox{2.5cm}{\diagAA}\hspace{12pt}$\DS=$\hspace{12pt}$\DS\frac{\mu^\epsilon}{2}\left(-\frac{2}{\epsilon}+ln\Delta+\mbox{\small finitos}\right)$\\
\hspace*{3cm}$\DS\frac{1}{4\pi}\del^a\bar{X}^\sigma\del_a\bar{X}^\rho\times\left[\frac{1}{2}G_{\rho\sigma,\mu\nu}G^{\mu\nu}-G_{\mu\rho,\nu}G_{\lambda\sigma,\xi}G^{\lambda\nu}G^{\mu\xi}\right],$\hfill\parbox{1.5cm}{\beq \eeq}\\
onde $\Delta=m^2+x(x+1)k^2$, e temos implicitamente uma integral $\int_0^1dx$ sobre o logar\'{\i}tmo (onde $x$ \'e o par\^ametro de Feynman). 

O termo acima nos dar\'a o contra-termo da m\'etrica. De acordo com \cite{hull}, podemos reescrever a express\~ao acima de forma a obter: 
\beq \label{eq.contraR}
\delta_{G_{\mu\nu}}=\frac{\mu^\epsilon}{4\pi\epsilon}\left(R_{\mu\nu}-\nabla_{(\mu}\Gamma_{\nu)}\right).
\eeq
Onde $\Gamma_\nu=g_{\nu\mu} g^{\rho\zeta}\Gamma_{\rho\zeta}^\mu$.
Dentro da a\c c\~ao efetiva, o termo da derivada, no segundo fator, pode ser integrado por partes, ou seja:
\beq
\nabla_{(\mu}\Gamma_{\nu)}\del_aX^\mu\del^aX^\mu=\nabla_\mu(\Gamma_\nu\del_aX^\mu\del^aX^\nu)-2\Gamma_\nu\nabla_\mu(\del^aX^\nu)\del_aX^\mu.
\eeq
O primeiro termo do segundo membro \'e uma diferencial total, e o segundo termo \'e proporcional \`as equa\c c\~oes de movimento, Eq.~(\ref{eq.sigmamov}), anulando-se na camada de massa. Logo o \'unico termo relevante para o contratermo \'e o tensor de Ricci
.

Podemos agora renormalizar a m\'etrica $G_{\mu\nu}$ definindo a m\'etrica nua como sendo:
\beq \label{eq.metricanua}
G_{\mu\nu}=\mu^\epsilon (G^R_{\mu\nu}+\frac{1}{4\pi\epsilon}R_{\mu\nu}).
\eeq

A fun\c c\~ao $\be^G$ da m\'etrica pode ser calculada, utilizando da equa\c c\~ao acima:
\beq \label{eq.betametrica}
\be^G_{\mu\nu}=\mu\frac{\del G^R_{\mu\nu}}{\del \mu}=-\epsilon\mu^{-\epsilon}G_{\mu\nu}=\frac{-1}{4\pi}R_{\mu\nu}.
\eeq


{\bf A Invari\^ancia Conforme.}
\\

A a\c c\~ao de Polyakov \'e classicamente invariante por transforma\c c\~oes conformes. Esta simetria reduz em muito o n\'umero de variedades inequivalentes a serem somadas e \'e uma caracter\'{\i}stica marcante da teoria de cordas\footnote{Veja \cite{green}, cap. 3.4.3}.
No n\'{\i}vel qu\^antico, por\'em, \'e necess\'ario verificar as circunst\^ancias nas quais a invari\^ancia conforme \'e mantida.

Seja a m\'etrica $\mr{g}_{\mu\nu}$, um escalar $\Lambda$ e vetores $k_\mu$ infinitesimais. Se para algum $\Lambda$:

$$\delta \mr{g}_{\mu\nu}=\nabla_\mu k_\nu+\nabla_\nu k_\mu+2\Lambda \mr{g}_{\mu\nu}=0,$$ 
ent\~ao $k_\mu$ s\~ao os vetores de Killing conformes. (se $\Lambda=0$, s\~ao apenas vetores de Killing).
Abaixo temos como algumas grandezas geom\'etricas transformam-se por transforma\c c\~ao conforme:
$$\mr{g}_{\mu\nu}\rightarrow e^{2\Lambda(x)}\mr{g}_{\mu\nu}, \qquad \sqrt{\mr{g}}\rightarrow \sqrt{\|e^{2\Lambda}\mr{g}_{\mu\nu}\|}=e^{n\Lambda(x)}\sqrt{\mr{g}},$$
$$\sqrt{\mr{g}}\mr{g}^{\mu\nu}\rightarrow e^{\epsilon\Lambda}\sqrt{\mr{g}}\mr{g}^{\mu\nu}.$$ 

A invari\^ancia por mudan\c cas de escala (transforma\c c\~ao conforme) est\'a relacionada com os zeros (pontos cr\'{\i}ticos) da fun\c c\~ao beta, uma vez que nestes pontos a constante de acoplamento n\~ao varia com a escala de energia (que \'e inversamente proporcional \`a escala de dist\^ancias). Para evidenciar esta rela\c c\~ao, consideremos apenas as transforma\c c\~oes conformes definamos um escalar que est\'a ligado ao tensor momento-energia\footnote{Como as transforma\c c\~oes conformes n\~ao t\^em dire\c c\~oes preferenciais, a grandeza conservada tamb\'em n\~a pode conter \'{\i}ndices, deve ser um escalar.}:
\beq
\delta S=\int d^n x\sqrt{\mr{g}}\Lambda(x)T.
\eeq

A partir desta defini\c c\~ao, calculemos $T$ explicitamente:
\beq
\delta S=\frac{\delta S}{\delta \mr{g}_{\mu\nu}}.\delta \mr{g}_{\mu\nu}=\frac{\delta S}{\delta \mr{g}_{\mu\nu}}.2\Lambda\mr{g}_{\mu\nu}=\int d^nx 2\Lambda \mr{g}_{\mu\nu}\frac{\delta S}{\delta \mr{g}_{\mu\nu}},
\eeq
implicando em
\beq
\sqrt{\mr{g}}T=2\mr{g}_{\mu\nu}\frac{\delta S}{\delta \mr{g}_{\mu\nu}}.
\eeq

Usando agora a a\c c\~ao (\ref{polyakov}), calculemos o termo acima explicitamente:
\beq\label{eq.LT}
2\mr{g}_{\mu\nu}\frac{\delta S}{\delta \mr{g}_{\mu\nu}}\rightarrow \epsilon L = T,
\eeq
onde neste c\'alculo deve-se atentar para os fatos de que $\delta \sqrt{\mr{g}}=\frac{1}{2}\sqrt{\mr{g}}\mr{g}^{\al\be}\delta\mr{g}_{\al\be}$, que $\frac{\delta \mr{g}^{\al\be}}{\delta \mr{g}_{\mu\nu}}=-\mr{g}^{\al\mu}\mr{g}^{\be\nu}$ e que $n=2+\epsilon$.


Utilizando a a\c c\~ao efetiva teremos:
\beq
\begin{array}{l}
\DS\delta\Gamma=e^\Gamma\int{\cal D}\Pi(-\delta S)e^{-S+\Pi.J}\nonumber\\
\DS\delta\Gamma=\left<\int d^2x\sqrt{ \mr{g}}\Lambda(x)T\right>=\int d^2x\sqrt{ \mr{g}}\Lambda(x)\left<T\right>
\end{array}
\eeq

Utilizando o resultado da Eq.(\ref{eq.LT}), teremos que:
\beq
T=\epsilon {\cal L}\rightarrow \epsilon\left<{\cal L}\right>.
\eeq

Esta \'e a {\bf anomalia conforme}. Observe que a priori $\left<T\right>$ deve ser zero, mas o valor esperado da Lagrangeana n\~ao \'e necessariamente zero. 
Integrando a anomalia, teremos:
\beq
\int d^2x\sqrt{ \mr{g}}\left<T\right>=\epsilon\left<S\right>.
\eeq

A a\c c\~ao depende linearmente da m\'etrica nua $G_{\mu\nu}$, ent\~ao podemos escrever:
\beq\label{eq.renormS}
\epsilon S=\epsilon G_{\mu\nu}.\frac{\delta S}{\delta G_{\mu\nu}}.
\eeq 

Fazendo uso da Eq.~(\ref{eq.metricanua}), que relaciona as m\'etricas nua e renormalizada, podemos ver que:

\beq
\DS\epsilon G=\mu\frac{\delta G}{\delta \mu}=\mu\frac{\delta G^R}{\delta \mu}\frac{\delta G}{\delta G^R}=\beta_G \frac{\delta G}{\delta G^R}.
\eeq

Relacionando esta equa\c c\~ao com a Eq.~(\ref{eq.renormS}), e calculando o valor esperado atrav\'es da a\c c\~ao efetiva, teremos:
\beq
\left<\epsilon S\right>=\left<\beta_G\frac{\delta G}{\delta G^R}\frac{\delta S}{\delta G}\right>=\beta_G\frac{\delta\Gamma}{\delta G^R}.
\eeq

A anomalia integrada fica da forma:
\beq
\int d^nx \sqrt{\mr{g}} \left<T\right>=\beta_G\frac{\delta \Gamma}{\delta G^R}.
\eeq
Para a teoria ser livre da anomalia conforme, devemos verificar o ponto em que a fun\c c\~ao beta da m\'etrica se anula, ou seja:
\beq
\beta_G\frac{\delta\Gamma}{\delta G^R}=0\rightarrow \be_G=0.
\eeq

Desta forma chagamos ao resultado de que, para a teoria de cordas manter a invari\^ancia conforme, o segundo fator da Eq.~(\ref{eq.contraR}) deve anular-se. O termo da derivada de $\Gamma$ anula-se, e a equa\c c\~ao que deve ser obedecida \'e:
\beq
R_{\mu\nu}=0,
\eeq
que \'e na verdade a {\it Equa\c c\~ao de Einstein da gravita\c c\~ao} para o v\'acuo.



\chapter{Supersimetria}
\section{Introdu\c{c}\~ao}

Simetrias de uma a\c{c}\~ao s\~ao definidas como transforma\c{c}\~oes nos campos da a\c{c}\~ao que deixam esta invariante. Os exemplos mais comuns s\~ao as invari\^ancias de Lorentz e gauge das teorias de Yang-Mills. Uma a\c{c}\~ao pode conter campos bos\^onicos, comutantes, e campos fermi\^onicos, anti-comutantes, ou grassmannianos. 

\'E poss\'{\i}vel construir transforma\c{c}\~oes que levam campos bos\^onicos para campos fermi\^onicos e vice-versa. Uma delas \'e a {\em Supersimetria}, que relaciona campos escalares, e campos espinoriais. Para definirmos esta simetria precisamos observar que em um espa\c co de $D$ dimens\~oes, os espinores de Weyl t\^em $2^{D/2}$ componentes reais. Podemos ent\~ao introduzir a supersimetria como: 

Em uma dimens\~ao: 
\beq
\delta_\epsilon \phi = \epsilon \psi,\qquad \delta \psi=-i\epsilon\del\phi.
\eeq


Em duas dimens\~oes:
\beq \label{eq.transsusy1}
\delta_\epsilon \phi = \epsilon^\al \psi_\al,\qquad \delta_\epsilon \psi_\be=-i\epsilon^\al \sigma_{\al\be}^m \del_m \phi,
\eeq
onde $\sigma_{\al\be}^m$ s\~ao as matrizes de Pauli em duas dimens\~oes e $\epsilon$, o par\^ametro da transforma\c{c}\~ao, \'e um espinor de Weyl\footnote{Supersimetria em quatro dimens\~oes pode ser encontrada em \cite{WessBagger}.}.

Utilizaremos as seguintes defini\c c\~oes para contra\c c\~ao e manipula\c c\~ao de \'{\i}ndices:
\begin{eqnarray}
\psi^\al&=&\epsilon^{\al\be}\psi_\be,\\
\psi_\al&=&\epsilon_{\al\be}\psi^\be,\\
\phi\sigma^m\xi&=&\phi^\al \sigma^m_{\al\be}\xi^\be,
\end{eqnarray}
onde $\epsilon_{\al\be}$ \'e o tensor completamente antissim\'etrico. 

\'E usual definir, em duas dimens\~oes, as matrizes $\sigma$ de Pauli como sendo
\beq
\sigma^0_{\al\be}=
		\left(\begin{array}{rr} 1&0\\0&1 \end{array}\right),
\sigma^1_{\al\be}= 
		\left(\begin{array}{rr} 1&0\\0&-1 \end{array}\right),
\quad \mbox{tal que} \quad 
\begin{array}{l}\sigma^0_\al{}^\be=
		\epsilon_{\al\be}\sigma^0_{\al\sigma}=
			\left(\begin{array}{rr}0&-1\\1&0 \end{array} \right)\\
\sigma^1_\al{}^\be=
		\epsilon_{\al\be}\sigma^1_{\al\sigma}=
			\left(\begin{array}{rr}0&+1\\1&0 \end{array} \right). \end{array} 
\eeq

Neste trabalho utilizaremos as coordenadas de cone de luz, definidas como
\beq\label{eq.ccl}
x^+=\frac{1}{2}(x^0+x^1),\quad x^-=\frac{1}{2}(x^0-x^1),
\eeq
onde $x^0$ e $x^1$ s\~ao as coordenadas usuais de Minkovisky. Neste sistema de coordenadas, as transforma\c c\~oes de supersimetria, Eq.~(\ref{eq.transsusy1}), escrevem-se como:
\beq\label{eq.transsusy2}
\delta_\epsilon \phi = \epsilon^\al \psi_\al,\qquad \delta_\epsilon \psi_\pm=-i\epsilon^\pm \del_\pm \phi,
\eeq
uma vez que $\sigma^\pm_{\al\be}=\meio (\sigma^0_{\al\be}\pm\sigma^1_{\al\be})$.

Exemplifiquemos, com uma a\c c\~ao bem simples, a supersimetria. Consideremos a a\c c\~ao de um escalar real livre e de um espinor real, no cone de luz:
\beq \label{eq.susy1}
\s=\int d^2x (\phi \del_+\del_-\phi+i\psi_+\del_-\psi_++i\psi_-\del_+\psi_-).
\eeq
Para constatar sua invari\^ancia sob transforma\c{c}\~oes de supersimetria, calculemos a varia\c{c}\~ao da a\c{c}\~ao.
\begin{eqnarray}
\delta_\epsilon \s&=&\int d^2x (2\delta_\epsilon\phi \del_+\del_-\phi+2i\delta_\epsilon\psi_+\del_-\psi_++2i\delta_\epsilon\psi_-\del_+\psi_- )=\nonumber \\
&=&\int d^2x (2\epsilon^\al\psi_\al\del_+\del_-\phi+2\epsilon^+\del_+\phi\del_-\psi_++2\epsilon^-\del_-\phi\del_+\psi_-)= 0. 
\end{eqnarray}
Logo vemos que esta a\c c\~ao \'e de fato supersim\'etrica, onde utilizamos o fato de que $\epsilon^\al$ \'e uma constante, e integramos por partes ao final.

Para sermos mais completos, precisamos definir a \'algebra de supersimetria. Da mesma forma que rota\c c\~oes em um plano definem $SO(2)$, podemos escrever a supersimetria de forma que defina um grupo de Lie. Muito de um grupo de Lie \'e determinado pela sua \'algebra, ent\~ao, definamos a \'algebra de supersimetria:
\beq \label{eq.algebrasusy}
[\delta_\epsilon,\delta_\zeta]=-2\epsilon^\al\sigma^m_{\al\be}\zeta^\be P_m,
\eeq  
onde $\delta_\epsilon, \delta_\zeta$ s\~ao elementos da \'algebra (transforma\c c\~oes infinitesimais) e $P_m$ \'e o momento, $-i\del_m$. 
Esta \'algebra deve ser satisfeita quando atuada sobre os campos de uma a\c c\~ao supersim\'etrica. Testemo-la sobre o campo bos\^onico $\phi$.

\begin{eqnarray}
[\delta_\epsilon, \delta_\zeta ]\phi&=&\delta_\epsilon (\zeta^\al \psi_\al)-\delta_\zeta(\epsilon^\al\psi_\al)=\nonumber\\
&=& 2i\epsilon^\al \sigma^m_{\al\be} \zeta^\be\del_m \phi.
\end{eqnarray}
Vemos ent\~ao que para o campo $\phi$ a \'algebra \'e satisfeita. Por outro lado, encontraremos algo diferente ao aplicarmos a \'algebra sobre os campos fermi\^onicos, $\psi$. Observe que 
\begin{eqnarray}\label{eq.alpsi}
[\delta_\epsilon, \delta_\zeta ]\psi_\pm&=&\delta_\epsilon (-i\zeta^\pm\del_\pm\phi)-\delta_\zeta(-i\epsilon^\pm\del_\pm\phi)=\nonumber\\
&=&-i\zeta^\pm\epsilon^\al\del_\pm\psi_\al+i\epsilon^\pm\zeta^\al\del_\pm\psi_\al,
\end{eqnarray}
n\~ao \'e o segundo membro de (\ref{eq.transsusy2}), podemos, no entanto, fazer a \'algebra cumprir-se caso utilizemos as equa\c c\~oes de movimento do campo $\psi$, a saber,
\beq
\del_\pm\psi_\mp=0.
\eeq
Aplicando estas equa\c c\~oes no segundo membro de (\ref{eq.alpsi}), teremos:
\beq
-i\zeta^\pm\epsilon^\al\del_\pm\psi_\al+i\epsilon^\pm\zeta^\al\del_\pm\psi_\al= -i\zeta^\pm\epsilon^\pm\del_\pm\psi_\pm+i\epsilon^\pm\zeta^\pm\del_\pm\psi_\pm=2i\epsilon^\pm\zeta^\pm\del_\pm \psi_\pm,
\eeq
satisfazendo, ent\~ao, a \'algebra de supersimetria.
Logo vemos o fato interessante de que a \'algebra s\'o se cumpre na camada de massa dos campos $\psi$. Para contornar este fato e termos a \'algebra fechando fora da camada de massa, precisamos introduzir um campo auxiliar bos\^onico nas transforma\c c\~oes de supersimetria. Estas ficam:
\beq\label{eq.transsusy3}
\begin{array}{rcl}
\delta_\epsilon \phi&=&\epsilon^\al \psi_\al \\
\delta_\epsilon \psi_\be&=&-i\epsilon^\al \sigma_{\al\be}^m \del_m \phi -i \epsilon_\be F \\
\delta_\epsilon F&=&\epsilon^\al\sigma_{\al\be}^m\del_m\psi^\be.
\end{array}
\eeq
Os campos $\phi$, $\psi^\al$ e $F$ formam o {\em multipleto} supersim\'etrico escalar. As transforma\c{c}\~oes acima satisfazem a \'algebra de supersimetria, Eq.~(\ref{eq.algebrasusy}), sem nescessidade de equa\c{c}\~oes extras. Podemos incluir o campo $F$ na a\c c\~ao como um campo auxiliar, que pode ser eliminado algebricamente da a\c c\~ao. Teremos ent\~ao:
\beq
\s=\int d^2x (\phi \del_+\del_-\phi+i\psi_+\del_-\psi_++i\psi_-\del_+\psi_--FF).
\eeq

\section{Supercampos}

Para evidenciar a supersimetria em uma teoria, consideremos que o espa\c{c}o seja formado pelas $D$ dimens\~oes usuais, $x^\mu$, bos\^onicas, e por $2^{D/2}$ dimens\~oes extras por\'em, mapeadas por coordenadas {\bf grassmanianas}, $\theta^\al$, fermi\^onicas. As coordenadas f\'ermions s\~ao espinores de Weyl, como acima.
Desta forma podemos definir a ``supercoordenada'':
\beq
\X^M=(x^m, \theta^\al).
\eeq
O espa\c{c}o mapeado pela supercoordenada chama-se {\bf superespa\c{c}o}.

Antes de continuarmos, relembremos o seguinte fato. O espa\c co euclidiano mapeado pelas coordenadas $x^m$ pode ser visto como o espa\c co tangente, do grupo de transla\c c\~oes gerado pelo momento $P_m$, ou seja, a \'algebra. Desta forma, considerando um vetor $x^mP_m$, da \'algebra, o correspondente elemento do grupo das transla\c c\~oes \'e definido como:
\beq
\ell_x=e^{-ix^mP_m}.
\eeq
Este grupo obedece \`a \'algebra abeliana das transla\c c\~oes: $[P_m,P_n]=0$, o que implica em:
\beq
\ell_x\ell_y=e^{-ix^mP_m}e^{-iy^mP_m}=e^{-i(x+y)^m P_m},
\eeq
onde usamos (e usaremos a seguir) a rela\c c\~ao $e^Ae^B=e^{A+B+\frac{1}{2}[A,B]}$, se $[A,[A,B]]=0$, que \'e o nosso caso.

Estas transforma\c c\~oes entre elementos do grupo induzem uma transforma\c c\~ao nos vetores da \'algebra:
\beq
x^m\rightarrow (x+y)^m.
\eeq
Desta forma, $P_m$ pode ser escrito de maneira a gerar estas transforma\c c\~oes no espa\c co tangente, a saber: $P_m=-i\del_m$. Assim
\beq
e^{y^m\del_m}f(x)=(1+y^m\del_m+\meio y^my^n\del_m\del_n+...)f(x)=f(x+y),
\eeq
pois esta \'e a expans\~ao de Taylor de $f(x+y)$ em torno de $x$.
Retornando \`a supersimetria, definamos um gerador $Q_\al$, tal que $\delta_\xi=\xi^\al Q_\al$, e que satisfa\c ca a \'algebra de supersimetria Eq.~(\ref{eq.algebrasusy}). Devido ao fato de que $Q_\al$ \'e grassmaniano, a \'algebra a ser obedecida fica na forma:
\beq \label{eq.algebra1}
\{Q_\al,Q_\be\}=2\sigma_{\al\be}^m P_m. 
\eeq

Podemos pensar agora o superespa\c co como sendo a \'algebra do grupo de supersimetria, gerado por $P_m$ e $Q_\al$. Analisando a parte fermi\^onica, podemos definir o elemento de ``supertransla\c c\~ao'' como sendo:
\beq
\ell_\theta=e^{\theta^\al Q_\al},
\eeq
e como sua \'algebra n\~ao \'e abeliana, a composi\c c\~ao assume a forma:
\beq
\ell_\theta\ell_\xi=e^{\theta^\al Q_\al}e^{\xi^\al Q_\al}=e^{(\theta+\xi)^\al Q_\al+\theta \sigma^m\xi P_m},
\eeq
ou seja, a composi\c c\~ao de duas transla\c c\~oes fermi\^onicas gera uma transla\c c\~ao espacial bos\^onica: 
\beq
\theta\sigma^m\xi P_m=-i\delta x^\mu P_m\rightarrow \delta x^m=i\theta \sigma^m \xi.
\eeq

Definindo uma base coordenada $\del_\al=\frac{\del}{\del \theta^\al}$ e $\del_m$, podemos definir $Q_\al$ como:
\beq
Q_\al=\del_\al-i\theta^\be \sigma^m_{\al\be}\del_m,
\eeq
este operador induzir\'a a transforma\c c\~ao no espa\c co tangente:
\beq
(1+\xi^\al Q_\al)\X=(1+\xi^\al Q_\al)(x^m+\theta^\be)=(x^m-i\theta\sigma^m\xi)+(\theta+\xi)^\be.
\eeq

Define-se usualmente a derivada covariante supersim\'etrica como sendo:
\beq
D_\al=\del_\al+i\theta^\be\sigma^m_{\be\al}\del_m.
\eeq
Este operador anticomuta com o gerador de supersimetria e satisfaz uma \'algebra muito semelhante \`a (\ref{eq.algebra1}), por\'em com o sinal trocado:
\begin{eqnarray}
\{D_\al,D_\be\}&=&-2\sigma_{\al\be}^m P_m,\\
\{D_\al,Q_\be\}&=&0.
\end{eqnarray}

Observe que \'e de extrema import\^ancia que a derivada covariante anticomute com o gerador de supersimetrias, pois isso nos diz que $D_\al$ \'e supersim\'etrica. 

Podemos agora definir os supercampos. Supercampos s\~ao fun\c c\~oes que dependem de \X: $f(\X)=f(x,\theta)$. Como as vari\'aveis grassmanianas anticomutam, uma expans\~ao em pot\^encias de $\theta^\al$ s\'o se desenvolver\'a at\'e o termo $\theta^{2^{D/2}}$. No caso bidimensional (n=2), teremos que:
\beq
f(x,\theta)=f(x, \theta)|_{\theta=0}+\theta^\al\del_\al f(\X)|_{\theta=0}-\frac{1}{4}\theta^2 \del^\al\del_\al f(\X)|_{\theta=0}.
\eeq
\'E usual definir campos dependentes apenas de $x^m$ para cada termo da expans\~ao. Ent\~ao, a  maneira comum de se definir a expans\~ao de um supercampo em termo dos seus campos componentes \'e:
\beq\label{eq.componentes}
f(x,\theta)=\phi(x)+\theta^\al\psi_\al(x)+\frac{1}{2}\theta^2F(x).
\eeq
Assim evidenciamos os campos componentes do supercampo, a saber, um campo escalar, um campo espinorial e um \'ultimo campo escalar.

\'E importante ver que uma lagrangeana definida a partir de supercampos, \'e supersim\'etrica.

\beq
\int Q_\al L(\X)d^2x d^2\theta =\int (\del_\al+i\theta^\be \sigma_{\be\al}^m \del_m) L(\X) d^2x d^2\theta=0,
\eeq
pois, o primeiro termo \'e uma derivada total, e o segundo pode ser integrado por partes.      

Em duas dimens\~oes e no cone de luz, podemos definir o gerados de supersimetria como:
\begin{eqnarray}
Q_+=\frac{\del}{\del \theta^+}-i\theta^+\partial_+ \\
Q_-=\frac{\del}{\del \theta^-}-i\theta^-\partial_-,
\end{eqnarray}
que satisfaz a \'algebra usual: 
\beq
\left\{ Q_{+ \atop -}, Q_{+ \atop -}\right\}=2P_{+ \atop -}
\eeq
e zero nos outros casos.
Da mesma forma, as derivadas covariantes escrevem-se:
\begin{eqnarray}
D_+=\frac{\del}{\del \theta^+}+i\theta^+\partial_+ \\
D_-=\frac{\del}{\del \theta^-}+i\theta^-\partial_-
\end{eqnarray}

Claramente podemos reescrever estas express\~oes utilizado das matrizes $\sigma$: $D_\al=\frac{\del}{\del\theta^\al}+i\theta^\be\sigma^m_{\al\be}\del_m$.

\section{M\'ultiplas supersimetrias}
Assim como definimos a transforma\c{c}\~ao de Supersimetria entre um campo escalar e um conjunto de f\'ermions, podemos definir mais simetrias semelhantes relacionando o mesmo campo bos\^onico com outros campos fermi\^onicos, ou seja, mais supersimetrias. Chamaremos o n\'umero de supersimetrias de N. N=2, duas supersimetrias, N=1, uma, N=0 \'e o caso bos\^onico.  

Para implementar estas novas simetrias no formalismo j\'a exposto, consideremos v\'arios geradores de supersimetrias: $Q^A_\al$, onde $A$ indica qual gerador estamos nos referindo. A \'algebra de Susi assume a forma:
\beq
\{Q_\al^A, Q_\be^B\}=2\sigma^m_{\al\be}P_m M^{AB}.
\eeq
Consideraremos $M^{AB}=I^{AB}$ no nosso caso\footnote{Este $M$ est\'a relacionado com a carga central definida em \cite{WessBagger}}. 
Seguindo, podemos associar a cada nova supersimetria um novo conjunto de vari\'aveis fermi\^onicas $\theta^A_\al$, e definir supercampos com N supersimetrias. Como exemplo, fa\c camos o caso N=2 em duas dimens\~oes:
\beq
f(x, \theta, \xi)=f(x)+\theta^\al\psi_\al+\xi^\al\chi_\al+\theta^\al\xi^\be\Xi_{\al\be}+\meio\theta^2F+\meio\xi^2G+\meio\theta^2\xi^\al\Sigma_\al+\meio\xi^2\theta^\al\Theta_\al+\frac{1}{4}\xi^2\theta^2A,
\eeq
onde os campos do multipleto acima v\^em diretamente da expans\~ao da Taylor de $f(x, \theta, \xi)$ em torno dos pontos $\theta$ e $\xi$.


Podemos definir a\c c\~oes com m\'ultiplas supersimetrias da mesma forma como o caso N=1, assim teremos
\begin{eqnarray}
\s&=&\int d^nx \prod^N d^{2^{D/2}}\theta^A L(x,\theta^A)\mbox{, ou no caso N=2,}\nonumber\\
\s&=&\int d^2x d^2\theta d^2\xi L(x,\theta, \xi).
\end{eqnarray}
Este \'e o formalismo manifestamente supersim\'etrico.

\vskip .5cm

Existem outras maneiras de termos supersimetrias extras que n\~ao apresentam-se de maneira manifesta. Uma delas pode ser definida da seguinte maneira. Seja uma a\c c\~ao com supersimetria N=1 manifesta contendo v\'arios supercampos escalares $\phi^i$, onde o \'{\i}ndice indica qual campo estamos nos referindo. Outras supersimetrias podem ser introduzidas se considerarmos a seguinte transforma\c c\~ao: 
\beq\label{eq.multiplas}
\delta_\epsilon \phi^i = \epsilon^\al J^i{}_j D_\al \phi^j.
\eeq
Podemos considerar estas equa\c c\~oes como sendo varia\c c\~os das Eqs.~(\ref{eq.transsusy1}), mas para serem supersimetrias, a \'algebra (\ref{eq.algebrasusy}) deve ser satisfeira. 
\begin{eqnarray}
[\delta_\epsilon,\delta_\xi]\phi^i&=&\delta_\epsilon(\xi^\al J^i{}_j D_\al\phi^j)-\delta_\xi(\epsilon^\al J^i{}_j D_\al \phi^j)=\nonumber\\
&=&\xi^\al J^i{}_{j,k}\delta_\epsilon \phi^kD_\al\phi^j+\xi^\al J^i{}_jD_\al(\delta_\epsilon \phi^j)-\quad (\epsilon \leftrightarrow \xi)=\nonumber\\
&=&-\epsilon^\al\xi^\be J^i{}_jJ^j{}_l \left\{D_\al,D_\be\right\}\phi^l+\xi^\al\epsilon^\be\left(J^i{}_{j,k}J^k{}_l-J^i{}_{l,k}J^k{}_j-\right.\nonumber\\
&&\left.{}-J^i{}_jJ^j{}_{j,k}+J^i{}_jJ^j{}_{k,l}\right)D_\be\phi^lD_\al\phi^j.
\end{eqnarray}
Para que tenhamos 
\beq
[\delta_\epsilon,\delta_\xi]\phi^i=2i\epsilon^\al\xi^\be\sigma^m_{\al\be}\del_m\phi^i,
\eeq
teremos que ter o termo dos anticomutadores multiplicado por $-1$, e o restante nulo. Isto implica nas condi\c c\~oes \cite{Spindel,Sevrin,Jack}:
\beq
J^i{}_jJ^j{}_k=-\delta^i{}_k,\mbox{ e }J^l{}_iJ^k{}_{[j,l]}-J^l{}_jJ^k{}_{[i,l]}=0.
\eeq

Estas condi\c c\~oes coincidentemente s\~ao as condi\c c\~oes suficientes para que $J$ seja uma estrutura complexa, implicando que $\phi^j$ sejam coordenadas de uma variedade de K\"ahler. Isto ser\'a melhor explicado na se\c c\~ao \ref{sec.kahler}. De qualquer maneira \'e importante ressaltar que para termos mais supersimetrias com o mesmo conte\'udo de campos, o espa\c co alvo tem que ser K\"ahler.


\section{Duas Supersimetrias}
Al\'em das coordenadas bos\^onicas $x^m$ e das duas dire\c{c}\~oes fermi\^onicas $\theta^\al (\equiv \theta^\al_1)$, introduzamos mais duas dire\c{c}\~oes, $\theta_2^\al$, com derivadas covariantes e geradores de supersimetria id\^enticos aos mesmos para as vari\'aveis $\theta_1^\al$. 

Assim teremos:
\beq
Q_{A\al} = \del_{\theta_A^\al}-i\theta_{A}^\be \sigma^m_{\al\be}\del_m,\qquad D_{A\al}= \del_{\theta_A^\al}+i\theta_{A}^\be \sigma^m_{\al\be}\del_m,
\eeq 
onde $\al,\be=1,2$, s\~ao os \'{\i}ndices espinoriais e $A,B=1,2$ s\~ao os \'{\i}ndices de supersimetria. Estes operadores ssatisfazem a \'algebra usual de supersimetria, Eq.(\ref{eq.algebra1}).

Agora podemos redefinir as vari\'aveis fermi\^onicas de uma forma mais conveniente:
\beq \begin{array}{l} \label{eq.DN=2}
\vartheta^\al=\frac{1}{\sqrt{2}}\left(\theta_1^\al-i\theta_2^\al\right),\qquad \btheta^\al=\frac{1}{\sqrt{2}}\left(\theta_1^\al+i\theta_2^\al\right). \\
\D^\al=\frac{1}{\sqrt{2}}\left(D_1^\al+iD_2^\al\right),\qquad \bar{\D}^\al=\frac{1}{\sqrt{2}}\left(D_1^\al-iD_2^\al\right).
\end{array}
\eeq
Que satisfaz a \'algebra:
\beq
\left\{\D_\pm,\bar{\D}_\pm\right\}=2P_\pm,
\eeq
onde aqui n\~ao h\'a soma sobre os \'{\i}ndices $\al$ e j\'a fizemos a escolha da representa\c c\~ao das matrizes sigma (coordenada de cone de luz, Eq.~(\ref{eq.ccl})).

Para construirmos uma a\c{c}\~ao manifestamente supersim\'etrica N=2, em duas dimens\~oes, teremos que considerar inicialmente que a medida da integral da a\c{c}\~ao \'e $d^2xd^2\theta_1d^2\theta_2$ cuja dimens\~ao de massa \'e nula, uma vez que $[d^2x]=-2$, $[d^2\theta]=-1$. Isto implica que qualquer lagrangeana proposta deve ser adimensional. Em variedades de grupos, a lagrangeana \'e expressa em termos de campos e seus inversos, logo n\~ao contribuindo para a dimens\~ao, desta forma a presen\c{c}a de derivadas n\~ao \'e permitida j\'a que estas contribuem para a dimens\~ao de massa. 

Desta forma, a\c{c}\~oes manifestamente supersim\'etricas N=2 em duas dimens\~oes t\^em que ser da forma:
\beq \label{eq.Kn=2}
\s=\int d^2x d^4\theta K(\vartheta_\al,\btheta_\al)=\int d^2x d^4\theta K(\theta_{1\al},\theta_{2\al}),
\eeq
onde $K$ \'e um campo escalar e $d^4\theta=d^2\theta_1 d^2\theta_2$. 

Como na a\c{c}\~ao (\ref{eq.Kn=2}) n\~ao existem derivadas, toda a din\^amica da teoria \'e definida por poss\'{\i}veis v\'{\i}nculos a serem satisfeitos pelo campo escalar $K$. \'E demonstrado que uma grande classe de modelos \sig podem ser compostos por campos quirais e quirais torcidos \cite{Sevrin}. Desta forma, restringir-nos-emos a potenciais da forma:
\beq \label{eq.Kquiral}
K(\theta_{1\al},\theta_{2\al})=K(\Phi,\bar{\Phi},\Lambda,\bar{\Lambda}).
\eeq

Definiremos a seguir estes dois tipos de campos:

\subsection{Campo Quiral e Quiral Torcido.}

Seja um supercampo N=2, $\Phi=\Phi(x, \theta_1, \theta_2)$, e seu conjugado complexo. Considerando que as quatro derivadas covariantes $\D_\al, \bar{\D}_\al$ definem, cada uma, dire\c c\~oes no superespa\c co, os campos quirais s\~ao definidos como aqueles que se propagam apenas na dire\c c\~ao holom\'orfica, ou seja:

\beq
\DB_\al\Phi=0,\qquad \D_\al\bar{\Phi}=0,
\eeq
onde a segunda express\~ao vem do fato dos campos conjugados (anti-quirais) propagarem-se na dire\c c\~ao anti-holom\'orfica.
Pela defini\c c\~ao de $\D_\al$, Eq.~(\ref{eq.DN=2}), estas condi\c c\~oes implicam que: 
\beq\label{eq.quiral1}
D_{1\al}\Phi=iD_{2\al}\Phi, \qquad D_{1\al}\bar{\Phi}=-iD_{2\al}\bar{\Phi},
\eeq
e tamb\'em temos que:
\beq\label{eq.quiral2}
D^2_2\Phi=-D^2_1\Phi,\qquad D^2_2\bar{\Phi}=-D^2_1\bar{\Phi}.
\eeq
De posse destas rela\c c\~oes cinem\'aticas, podemos encontrar rela\c c\~oes entre os componentes do multipleto $\Phi$. A expans\~ao em pot\^encias de $\theta_2$ nos fornece
\beq
\Phi(\theta_1,\theta_2)=\phi(\theta_1)+\theta^\al_2\psi_\al(\theta_1)+\theta_2^2F(\theta_1),
\eeq
em que aqu\'{\i} cada campo componente \'e um supercampo N=1.
Observe que por defini\c c\~ao,
\beq
\psi_\al=-iD_{2\al}\Phi|_{\theta_2}=-iD_{1\al}\phi,
\eeq
\beq F=-\frac{1}{4}D^2_2\Phi|_{\theta_2=0}=-\frac{1}{4}D_1^2\phi,
\eeq
onde usamos as Eq.~(\ref{eq.quiral1}) e Eq.~(\ref{eq.quiral2}).  Conclu\'{\i}mos que o campo quiral N=2 $\Phi$ tem apenas um campo independente no seu multipleto, a saber o campo N=1 $\phi$ que n\~ao satisfaz nenhum v\'{\i}nculo.

Repetindo os passos anteriores agora para o campo anti-quiral $\bar{\Phi}$, teremos que

\beq
\bar{\Phi}(\theta_1,\theta_2)=\bar{\phi}(\theta_1)+\theta^\al_2\bar{\psi}_\al(\theta_1)+\frac{1}{4}\theta_2^2\bar{F}(\theta_1),
\eeq
e a expans\~ao em campos componentes 
\beq\label{eq.transquiral}
\bar{\psi}_\al=iD_{2\al}\bar{\Phi}|_{\theta_2}=-iD_{1\al}\bar{\phi},
\eeq
\beq \bar{F}=-\frac{1}{4}D^2_2\bar{\Phi}|_{\theta_2=0}=-\frac{1}{4}D_1^2\bar{\phi}.
\eeq

Introduzamos agora os campos quirais torcidos (do ingl\^es {\it twisted chiral}), $\Lambda(x,\theta_1,\theta_2)$. Estes campos tamb\'em obedecem rela\c c\~oes cinem\'aticas por\'em diferentes das satisfeitas pelos campos quirais. Elas s\~ao

\beq
\DB_+\Lambda=\D_-\Lambda=0;\qquad \D_+\bar{\Lambda}=\DB_-\bar{\Lambda}=0.
\eeq

O t\'{\i}tulo {\it torcido} pode ser sugerido se observarmos que estes s\~ao campos quirais usuais onde trocamos $\bar{\D}_- \leftrightarrow \D_-$.

Estas condi\c{c}\~oes implicam que:

\beq
D_{+1}\Lambda=iD_{+2}\Lambda,\qquad D_{-1}\Lambda=-iD_{-2}\Lambda.
\eeq
\beq
D_{+1}\bar{\Lambda}=-iD_{+2}\bar{\Lambda},\qquad D_{-1}\bar{\Lambda}=iD_{-2}\bar{\Lambda},
\eeq

ou melhor:

\beq
D_{2\al}\Lambda=M_\al{}^\be D_{1\be}\Lambda, \qquad D_{2\al}\bar{\Lambda}=\bar{M}_\al{}^\be D_{1\be}\bar{\Lambda},
\eeq

onde,
\beq \label{eq.M} 
M_\al{}^\be=\left(\begin{array}{rr}-i&0\\0&i\end{array}\right), \qquad \bar{M}=M^*.
\eeq

Al\'em disto:
\beq 
D^2_2\Lambda=D^2_1\Lambda,\qquad D^2_2\bar{\Lambda}=D^2_1\bar{\Lambda}.
\eeq

Sob estas condi\c{c}\~oes, a expans\~ao de Taylor em torno de $\theta_2$ assume a forma:
\beq
\Lambda(\theta_1,\theta_2)=\lambda(\theta_1)+\theta^\al_2\chi_\al(\theta_1)+\theta^2_2 F(\theta_1),
\eeq onde,
\beq\label{eq.transtwisted}
\chi_\al=D_{2\al}\Lambda|_{\theta_2}=\left\{\begin{array}{r} -i D_{+1}\lambda \\
iD_{-1}\lambda
\end{array}\right.,
\eeq
\beq F=\frac{1}{4}D^2_2\Lambda|_{\theta_2=0}=\frac{1}{4}D^2_1\lambda.
\eeq

Novamente vemos que o campo N=2 quiral torcido $\Lambda$ depende apenas de sua primeira componente N=1, $\lambda$, que por sua vez n\~ao obedece nenhum v\'{\i}nculo.

Analogamente para o campo anti-quiral torcido (o conjugado complexo)
\beq
\bar{\Lambda}(\theta_1,\theta_2)=\bar{\lambda}(\theta_1)+\theta^\al_2\bar{\chi}_\al(\theta)+\theta^2_2 \bar{F}(\theta_1),
\eeq onde,
\beq
\bar{\chi}_\al=D_{2\al}\bar{\Lambda}|_{\theta_2}=\left\{\begin{array}{r} i D_{+1}\bar{\lambda} \nonumber \\
-iD_{-1}\bar{\lambda}
\end{array}\right. ,
\eeq
\beq \bar{F}=\frac{1}{4}D^2_2\bar{\Lambda}|_{\theta_2=0}=\frac{1}{4}\bar{\lambda}.
\eeq

O fato das componentes dos supercampos N=2 quiral e quiral torcido estarem relacionada via uma opera\c c\~ao determinada, Eq.~(\ref{eq.transquiral}) para as componentes do supercampo quiral e Eq.~(\ref{eq.transtwisted}), nos mostra que estas transforma\c c\~oes formam uma outra representa\c c\~ao da \'algebra de supersimetria. Chamando $\tilde{\Q}$ os novos geradores de superdimetria, podemos escreve-los explicitamente como 


\beq\label{eq.Q1}
\tilde{\Q}_\al\phi=-iD_{\al 1}\phi,\qquad \tilde{\Q}_\al\bar{\phi}=iD_{\al 1}\bar{\phi},\eeq

\beq \label{eq.Q2}
\tilde{\Q}_\al\lambda=M_\al{}^\be D_{\be 1}\lambda,\qquad \tilde{\Q}_\al\bar{\lambda}=\bar{M}_\al{}^\be D_{\be 1}\bar{\lambda}.
\eeq

Podemos escrever estas equa\c c\~oes de uma forma muito mais \'util para a seq\"u\^encia do trabalho. Definamos um vetor $\V\;^i$ como
\beq
\V\;^i=(\Phi, \bar{\Phi}, \Lambda, \bar{\Lambda}),
\eeq
assim definamos duas matrizes $J_+{}^i{}_j$ e $J_-{}^i{}_j$ que satisfazem
\beq
\tilde{\Q}_\al \V\;^i=J_\al{}^i{}_jD_{1\al}\V\;^j,
\eeq
observe que esta equa\c c\~ao \'e exatamente a Eq.~(\ref{eq.multiplas}). Atrav\'es da equa\c c\~oes (\ref{eq.Q1}) e (\ref{eq.Q2}), podemos determinar as matrizes $J$ e obteremos:
\beq\label{eq.Jestrutura}
J_+=diag(-i,i,-i,i),\mbox{ e, }J_-=diag(-i,i,i,-i).
\eeq

Voltando para a a\c c\~ao N=2, Eq.~(\ref{eq.Kn=2}), devemos utilizar os v\'{\i}nculos dos campos, apresentados acima, para que possamos integrar sobre as vari\'aveirs $\theta^\al_2$. Expandindo a a\c c\~ao (\ref{eq.Kn=2}), teremos:

\begin{eqnarray}\label{eq.Kn=1}
\s&=&\int d^2x d^4\theta K(\Phi,\bar{\Phi},\Lambda,\bar{\Lambda})=\frac{1}{4}\int d^2x d^2\theta \left[4K_{\phi\bar{\phi}}D^\al_1\phi D_{1\al}\bar{\phi}-4K_{\lambda\bar{\lambda}}D^\al_1\lambda D_{1\al}\bar{\lambda}+\right.\nonumber \\
&-&\left.2iK_{\phi\bar{\lambda}}D^\al_1\phi\bar{M}_\al{}^\be D_{1\be}\bar{\lambda}-2iK_{\phi\lambda}D^\al_1\phi M_\al{}^\be D_{1\be}\lambda\right.+\nonumber\\
&+&\left.2iK_{\bar{\phi}\lambda}D^\al_1\bar{\phi}M_\al{}^\be D_{1\be}\lambda+2iK_{\bar{\phi}\bar{\lambda}}D^\al_1\bar{\phi}\bar{M}_\al{}^\be D_{1\be}\bar{\lambda}\right],
\end{eqnarray}
onde os coeficientes significam, por exemplo:
\beq
K_{\Phi\Lambda}=\frac{\del^2 K}{\del\Phi\del\Lambda}.
\eeq
O que ent\~ao determinar\'a a f\'{\i}sica do modelo ser\'a a forma do potencial $K$.  

Esta express\~ao pode ser posta em uma forma bem mais conveniente e usual. Observemos que:
\beq
iM^{\al\be}=i\epsilon^{\al\gamma}M_\gamma{}^\be=i\left(\begin{array}{cc} 0&i\\i&0\end{array}\right)=
-\left(\begin{array}{cc} 0&1\\1&0\end{array}\right)\equiv\tilde{\delta}^{\al\be}.
\eeq
e desta forma, teremos que
\begin{eqnarray} \label{eq.Kn=12}
\s&=&\int d^2x d^2\theta \bigg[K_{\phi\bar{\phi}}D^\al_1\phi D_{1\al}\bar{\phi}-K_{\lambda\bar{\lambda}}D^\al_1\lambda D_{1\al}\bar{\lambda}+\nonumber \\
&+&\left(\frac{1}{2}K_{\phi\bar{\lambda}}D_{1\al} \phi D_{1\be}\bar{\lambda}-\frac{1}{2}K_{\phi\lambda}D_{1\al}\phi D_{1\be}\lambda\right.+\nonumber\\
&+&\left.\left.\frac{1}{2}K_{\bar{\phi}\lambda}D_{1\al}\bar{\phi}D_{1\be}\lambda-\frac{1}{2}K_{\bar{\phi}\bar{\lambda}}D_{1\al}\bar{\phi} D_{1\be}\bar{\lambda}\right)\tilde{\delta}^{\al\be}\right].
\end{eqnarray}

A a\c{c}\~ao mais simples N=2 \'e a de campos quirais livres. Esta tamb\'em pode ser chamada de modelo sigma linear, uma vez que os campos est\~ao relacionados de uma forma linear. A a\c c\~ao \'e da forma:
\beq
S=\int d^2x d^4\theta \bar{\Phi}\Phi=\int d^2x d^2\theta D_1^\al\phi D_{1\al} \bar{\phi}.
\eeq

Esta a\c{c}\~ao reduz-se, ap\'os o supercampo N=1 ser expandido por meios de  
\beq
\phi=A+\theta_1^\al B_\al+\frac{1}{2}\theta_1^2 C,
\eeq
e ser integrado em $\theta_1$, \`a express\~ao:
\beq
S=\int d^2x \left(\bar{A}\del_+\del_-A+i\bar{B}_+\del_-B_++i\bar{B}_-\del_+B_--\frac{1}{2}\bar{C}C\right).
\eeq
 
Onde temos um campo $A$ escalar complexo e um campo de spin $\frac{1}{2}$, $B$, ambos livres. O campo auxiliar $C$ pode ser retirado algebricamente atrav\'es de suas equa\c c\~oes de movimento
~$C=\bar{C}=0$.

Poder\'{\i}amos ter feito o mesmo para os campos quirais torcidos livres, por\'em, a a\c c\~ao seria a mesma, a menos de um sinal, que pode ser visto do segundo termo da Eq.~(\ref{eq.Kn=12}).



\chapter{Modelo Sigma Supersim\'etrico.}
\section{Uma Supersimetria.}

Como mencionado no cap\'{\i}tulo \ref{cap.sigmabos}, o modelo \sig  tamb\'em serve como prot\'otipo na cons\-tru\-\c{c}\~ao da teoria de cordas, ou em outras palavras, a teoria de corda bos\^onica \'e um tipo de modelo \sig em duas dimens\~oes onde temos a presen\c{c}a do gr\'aviton (termo principal), do campo antissim\'etrico (termo de WZW) e do d\'{\i}laton (que n\~ao foi exposto neste trabalho). Uma motiva\c{c}\~ao para se estudar uma generaliza\c{c}\~ao supersim\'etrica do modelo \sig \'e entender a teoria de supercordas. 


Existem duas formas de supersimetrizar a a\c{c}\~ao~(\ref{eq.sigmagrupo}). Uma delas \'e impor leis de transforma\c{c}\~ao semelhantes \`as equa\c{c}\~oes (\ref{eq.transsusy1}) e, a partir do termo bos\^onico, encontrar a contraparte fermi\^onica. Este m\'etodo \'e seguido em \cite{Spindel} mas a a\c{c}\~ao obtida \'e supersim\'etrica apenas na camada de massa, como a a\c{c}\~ao (\ref{eq.susy1}). 

Um m\'etodo mais f\'acil e direto \'e impor que os campos $\mr{g}$ e $\mr{b}$ n\~ao sejam fun\c{c}\~oes do campo $\phi(x)$ bos\^onico mas sim de um supercampo N=1 $\Phi(x,\mbox{ }\theta)$. Ent\~ao teremos:
\beq\label{eq.sigmasusy1}
\s=\int d^2xd^2\theta\frac{1}{2}D_-\Phi^iD_+\Phi^j(\mr{g}_{ij}+\mr{b}_{ij})(\Phi).
\eeq
Utilizamos as derivadas covariantes pois estas s\~ao as derivadas naturais que atuando sobre supercampos,  reduzem-se \`as derivadas usuais quando $\theta=0$. Veja que a a\c{c}\~ao est\'a dimensionalmente correta uma vez que $[D]=\frac{-1}{2}$, $[d\theta]=\frac{-1}{2}$ e $[dx]=1$.

Podemos calcular as equa\c{c}\~oes de movimento, que se escrevem como:
\beq
D_+D_-\Phi^i+\Gamma^i_{+jk}D_+\Phi^jD_-\Phi^k=0,
\eeq
onde 
\begin{eqnarray}
\Gamma^i_{\pm jk}&=&\mr{g}^{il}(\{l,jk\}\pm T_{ljk}) \\
\{l,jk\}&=&\frac{1}{2}(\mr{g}_{jl,k}+\mr{g}_{kl,j}-\mr{g}_{jk,l})\\
T_{ljk}&=&\mr{b}_{lj,k}+\mr{b}_{kl,j}+\mr{b}_{jk,l}=3\mr{b}_{[li,j]}.
\end{eqnarray}

Podemos escrever o supercampo $\Phi$ em termos de seus campos componentes, semelhante a expans\~ao (\ref{eq.componentes}), a saber:
\beq
\Phi^i(x,\theta)=\phi^i(x)+\theta^\al\psi^i{}_\al(x)_\al+\frac{1}{2}\theta^2F^i(x).
\eeq

Assim, a a\c{c}\~ao (\ref{eq.sigmasusy1}) pode ser reescrita como
\begin{eqnarray}
\s&=&\int d^2x \left\{-g_{jk}(i\psi^j_+\del_-\psi^k_+-F^jF^k+\del_+\phi^j\del_-\phi^k-i\del_+\psi^j_-\psi^k_-)-\right.\nonumber\\
&&{}-g_{jk,l}F^l\psi^j_+\psi^k_--\frac{1}{2}g_{jk,lm}\psi^{l\al}\psi^m_{\al}\psi^j_+\psi^k_-+ig_{jk,l}\psi^l_-\psi^k_-\del_+\phi^j-\nonumber\\
&&\left.-g_{jk,l}\psi^l_-\psi^j_+F^k+i g_{jk,l}\psi^l_+\psi^j_+\del_-\psi^k+g_{jk,l}\psi^l_+F^j\psi^k_-\right\}.
\end{eqnarray}  
Observe que esta a\c{c}\~ao cont\'em um campo auxiliar, $F$, n\~ao propagante. Este campo garante a supersimetria fora da camada de massa. Se desejarmos elimin\'a-lo, para obter a a\c{c}\~ao em termo apenas de campos f\'{\i}sicos, devemos substituir as equa\c{c}\~oes de movimento para $F$ diretamente na a\c{c}\~ao, a saber:
\beq \label{eq.movF}
F^i=-\Gamma^i_{+jk}\psi^i_+\psi^k_-,
\eeq
ent\~ao, ap\'os algum algebrismo, teremos:
\begin{eqnarray}
\s&=&\int d^2x \left[\frac{1}{2}\left(\mr{g}_{ij}\eta^{\mu\nu}+\mr{b}_{ij}\epsilon^{\mu\nu}\right)\del_\mu\phi^i\del_\nu\phi^j+\right.\nonumber\\
&&\left.\left[ i\mr{g}_{ij}\psi^i\dslash\psi^j+i\mr{g}_{ik}\Gamma^k_{+jl}\psi^l_-\psi^j_-\del_+\phi^i+\right.\right.\nonumber\\
&&\left.\left. i\mr{g}_{ik}\Gamma^k_{-jl}\psi^l_+\psi^j_+\del_-\phi^i\right]+\frac{1}{4}R_{ijkl}\psi^{i\al}\psi^k_\al\psi^{j\be}\psi^l_\be\right].
\end{eqnarray}

\vskip .5cm
Definamos o modelo sigma supersim\'etrico em variedades de grupos. Inicialmente devemos ter como base o desenvolvimento apresentado na se\c{c}\~ao \ref{sec.modgrupo}, onde o modelo \'e definido a partir de elementos $g$ do grupo {\got G}. A maneira mais natural de supersimetrizar os elementos de grupo \'e supersimetrizar sua \'algebra ou seja, ao inv\'es de termos vetores do tipo $\phi=\phi^i(x) \mb{T_i}$, teremos $\phi=\Phi^i \mb{T_i}$, onde $\Phi^j$ \'e um supercampo N=1, e $\mb{T_i}$ s\~ao os geradores do grupo. Como j\'a visto do caso bos\^onico, este supercampo \'e o mesmo que aparece na a\c{c}\~ao (\ref{eq.sigmasusy1}).
Via mapeamento exponencial podemos  definir os super-elementos N=1 de grupo como sendo:
\beq
\gs(x,\theta)=e^{\Phi^i\mb{T_i}}=g(x)+\theta^\al\psi_\al(x)+\frac{1}{2}\theta^2F(x).
\eeq
Observe que a primeira componente do multipleto \'e o elemento de grupo bos\^onico normalmente definido para grupos, $g(x)=\gs(x,\theta)|_{\theta=0}$.

Utilizando as rela\c{c}\~oes entre $\mr{g}$, $\mr{b}$ e $g$ definidas pelas Eq.~(\ref{eq.sigmaprincipal}) e Eq.~(\ref{eq.WZWH}), na se\c{c}\~ao (\ref{sec.tetrada}), onde agora quem descreve a teoria \'e um supercampo, teremos:
\begin{eqnarray}
\mr{g}_{ij}&=&\tr (\gs^{-1}_{,i}\gs_{,j})=-\tr (\gs^{-1}\gs_{,i}\gs^{-1}\gs_{,j}),\\
\del_{[i}\mr{b}_{jk]}&=&\frac{2}{3}\tr (\gs^{-1}\gs_{,[i}\gs^{-1}\gs_{,j}\gs^{-1}\gs_{,k]}).
\end{eqnarray}

A a\c{c}\~ao fica na forma
\begin{eqnarray}\label{eq.N=1}
\lefteqn{\s=\int d^2xd^2\theta \frac{1}{2}D_-\Phi^iD_+\Phi^j\tr\left(\gs^{-1}_{,i}\gs_{,j}\right)+}\nonumber\\
&&+\frac{2}{3}\int d^2xd^2\theta dt \frac{1}{2}D_-\Phi^iD_+\Phi^j\del_t\Phi^k\tr(\gs^{-1}\gs_{,[i}\gs^{-1}\gs_{,j}\gs^{-1}\gs_{,k]})=\nonumber\\
\lefteqn{\s=\int d^2xd^2\theta \tr(D^\al\gs^{-1}D_\al\gs)+}\nonumber\\
&&+\frac{2}{3}\int d^2xd^2\theta dt \delta^{\al\be}\tr(\gs^{-1}D_\al\gs\gs^{-1}D_\be\gs\gs^{-1}\del_t\gs).
\end{eqnarray}

Consideremos $\mr{b}=0$, ou seja, apenas o modelo principal, e calculemos as equa\c{c}\~oes de movimento. Expandindo o supercampo teremos
\begin{eqnarray}\label{eq.vinc}
\gs&=&g(x)+\theta^\al+\psi_\al+\meio\theta^2F,\nonumber\\
\gs^{-1}&=&g^{-1}+\theta^\al\tilde{\psi}_\al+\meio\theta^2\tilde{F}^{-1}. 
\end{eqnarray}

Como $\gs^{-1}\gs=1$, teremos $$g^{-1}g=1,\qquad \tilde{\psi}_\al=-g^{-1}\psi_\al g, \qquad \tilde{F}=-g^{-1}Fg^{-1}-g^{-1}\psi\tilde{\psi}.$$

Integrando nas vari\'aveis fermi\^onicas, teremos em componentes:

\beq
\int d^2x \left(-2g^{-1}\Box g +i\tilde{\psi}_-\lrarrow{\del_+}\psi_- +i\tilde{\psi}_+\lrarrow{\del_-}\psi_+ + 2 \tilde{F}F\right).
\eeq

Variando a a\c{c}\~ao, obtemos as equa\c{c}\~oes de movimento:
\beq
\del_m(\del^mg^{-1}g)=0,\quad \dslash\psi=0,\quad \del_m(g^{-1}\psi_\al g^{-1})\sigma^m=0,\quad F=g^{-1}\psi^\al g^{-1}\psi_\al g^{-1}=0.
\eeq

\section{Variedades de K\"ahler}\label{sec.kahler}

Uma variedade complexa \'e aquela cujos abertos podem ser mapeados para $\CC^n$ (o que signifca que podemos p\^or sobre a variedade coordenadas holom\'orficas, $(z^a)$, e anti-holom\'orficas, $\bar{z}^{\bar{a}}$) e cujas fun\c{c}\~oes de transi\c{c}\~ao entre os abertos sejam holom\'orficas e anti-holom\'orficas. ($z'^a=f^a(z)$, $\bar{z}'^{\bar{a}}=f^{\bar{a}} (\bar{z})$.)
Neste tipo de variedade existe uma transforma\c{c}\~ao que age em vetores:
\beq
v'^a=J_b{}^a v^b=iv^a,\qquad \bar{v}'^a=\bar{J}_b{}^a \bar{v}^b=-i\bar{v}^a.
\eeq
Esta transforma\c{c}\~ao \'e natural em variedades complexas. $J$ \'e chamado de estrutura complexa e \'e uma constante na variedade, uma vez que transforma\c{c}\~oes holom\'orficas de coordenadas entre abertos n\~ao alteram sua forma, pois
\beq   
J'^a_b=\frac{\del z'^a}{\del z^c}\frac{\del z^d}{\del z'^b}J^c_d=\frac{\del z'^a}{\del z^c}\frac{\del z^d}{\del z'^b} i \delta^c_d=i\delta^a_b.
\eeq
O mesmo vale para as transforma\c{c}\~oes anti-holom\'orficas.

Definamos ent\~ao as variedades de K\"ahler que s\~ao aquelas que admitem m\'etricas invariantes pelo grupo $U(N)$ de holonomia. Estas m\'etricas s\~ao chamadas de m\'etricas de K\"ahler \cite{green2}.
A holonomia $U(N)$ garante a possibilidade de se construir uma estrutura complexa pois $J$ \'e invariante por transforma\c{c}\~oes de $U(N)$ que s\~ao por defini\c{c}\~ao holom\'orficas. 

V\'arias variedades reais tamb\'em podem ser variedades complexas. De maneira geral, uma variedade real 2N-dimensional ter\'a como grupo de holonomia algum subgrupo do grupo $SO(2N)$. Se tivermos que o grupo de holonomia seja o grupo $U(N)$, ent\~ao a variedade ser\'a complexa. N\~ao \'e dif\'{\i}cil observar que grupo $SO(2N)$ cont\'em subgrupos $U(N)$, pois aquele divide-se em uma representa\c{c}\~ao holom\'orfica e uma representa\c{c}\~ao inequivalente anti-holom\'orfica. Para isso vejamos que vetores de $SO(2N)$ t\^em $2N$ componentes reais. Combinando suas componentes, duas a duas, podemos montar um vetor complexo de $N$ componentes. Ent\~ao podemos analisar como uma transforma\c{c}\~ao de $SO(2N)$ age nestes vetores complexos. Comparando com uma transforma\c{c}\~ao geral de $U(N)$, veremos que estas podem ser escritas como um caso particular daquelas.
Fazendo a mesma coisa com vetores conjugados complexos dos primeiros, teremos outra representa\c{c}\~ao de $U(N)$ que leva a um subgrupo de $SO(2N)$ diferente do primeiro. 
    
Assim, em variedades reais com grupo de holonomia $U(N)$ podemos fazer uma transforma\c{c}\~ao de coordenadas para coordenadas complexas e representar os vetores reais por dois vetores complexos, $A_i \rightarrow A_a, A_{\bar{a}}$. 

As m\'etricas de K\"ahler satisfazem, em coordenadas complexas:
\beq
\mr{g}_{a\bar{a}}=\mr{g}_{\bar{a}a},\quad \mr{g}_{aa}=\mr{g}_{\bar{a}\bar{a}}=0.
\eeq

Isto pode ser visto uma vez que as transforma\c c\~oes $U(N)$ deixam invariantes bilineares da forma $ds^2=\mr{g}_{a\bar{a}}dz^a d\bar{z}^{\bar{a}}$, onde $\mr{g}_{a\bar{a}}=A\delta_{a\bar{a}}$. 





A forma de K\"ahler \'e um objeto definido a partir da estrutura complexa:
\beq
{\cal K}_{ij}=\mr{g}_{ik}J^k{}_j.
\eeq
Em coordenadas holom\'orficas, onde $J=\pm i$, teremos:
\beq \label{eq.KG} \begin{array}{c}
{\cal K}_{a\bar{b}}=\mr{g}_{a\bar{c}}J^{\bar{c}}{}_{\bar{b}}=-i\mr{g}_{a\bar{b}}, \\
{\cal K}_{\bar{a}b}=\mr{g}_{\bar{a}c}J^c{}_b=i\mr{g}_{\bar{a}b}. \\
\end{array}
\eeq

 Observe que por constru\c{c}\~ao, ${\cal K}$ \'e um tensor covariante, antissim\'etrico e invariante por $U(N)$, pois \'e proporcional \`a m\'etrica.

Como ${\cal K}$ \'e por constru\c{c}\~ao uma constante, satisfaz:
\beq
\del_{[a}{\cal K}_{b]\bar{c}}=\del_{[\bar{a}}{\cal K}_{|b|\bar{c}]}=0. 
\eeq
Para satisfazer esta rela\c{c}\~ao, $K$ deve ser da forma:
\beq
{\cal K}_{a\bar{a}}=-i\del_{a}\del_{\bar{a}}K,
\eeq
onde $K$ \'e o chamado potencial de K\"ahler. Desta maneira podemos relacionar a m\'etrica com $K$, a saber:
\beq\label{eq.kahlermetrica}
\mr{g}_{a\bar{b}}=\frac{\del^2K}{\del z^a\del z^{\bar{b}}}.
\eeq

Este fato ser\'a de extrema import\^ancia para o caso do modelo sigma N=2, uma vez que a lagrangeana deste modelo \'e completamente determinada por um campo escalar.







\section{Modelo \sig no grupo $SU(2)\otimes U(1)$.}
Partamos agora para o exemplo final deste trabalho, onde usaremos os conceitos at\'e aqui desenvolvidos. 
J\'a sabemos que para termos mais de uma supersimetria precisamos de uma variedade complexa pois nela teremos as estruturas complexas. Para tanto, usaremos o grupo $SU(2)\otimes U(1)$ que na verdade \'e isom\'orfico ao grupo $U(2)$.

Podemos construir elementos deste grupo atrav\'es de duas vari\'aveis complexas, $a$, $b$ e seus conjugados complexos, $\bar{a}$, $\bar{b}$, da maneira:

\beq
g=\frac{e^{i \vartheta}}{d} \left(\begin{array}{rr}a&\bar{b}\\-b&\bar{a}\end{array}\right), \quad d=\sqrt{a\bar{a}+b\bar{b}},\quad \vartheta=-\frac{1}{2}\ln(d).
\eeq

Observe que sem o termo exponencial, uma fase, ter\'{\i}amos $g$ como elemento de $SU(2)$, por\'em com um grau de liberdade a mais. Esta fase faz a diferen\c ca entre o grupo $SU(2)$ e $U(2)$. 

Podemos construir uma representa\c{c}\~ao N=2 deste grupo, apresentada em \cite{Rocek}, que utiliza-se de campos quirais ($\Phi$) e quirais torcidos ($\Lambda$), a saber:

\beq \label{eq.elementoN=2}
g=e^{i\vartheta}(\Phi\bar{\Phi}+\Lambda\bar{\Lambda})^{-1/2}\left(\begin{array}{rr}\Lambda&\bar{\Phi}\\-\Phi&\bar{\Lambda}\end{array}\right),
\eeq
onde $\vartheta=-\frac{1}{2}\ln(\Phi\bar{\Phi}+\Lambda\bar{\Lambda})$.
Os campos $\Phi$, $\bar{\Phi}$, $\Lambda$ e $\bar{\Lambda}$ representam as coordenadas complexas da variedade de $SU(2)\otimes U(1)$. Desta forma, a estrutura complexa escreve-se diretamente como $\pm i$, fato este que pode ser visto da Eq.~(\ref{eq.Jestrutura}). Assim, ao utilisarmos estes campos j\'a garantimos a presen\c ca da estrutura complexa. 

A m\'etrica da variedade de grupo \'e definida como:
\beq
\mr{g}_{ij}=\tr(g^{-1}_{,i}g_{,j})=-\tr (g^{-1}g_{,i}g^{-1}g_{,j})
\eeq
onde $i$ e $j$ representam $\Phi,\bar{\Phi},\Lambda,\bar{\Lambda}$.

Calculando explicitamente as componentes da m\'etrica, obteremos que:
\beq 
\left\{\begin{array}{l}
\mr{g}_{\Phi\bar{\Phi}}=\mr{g}_{\Lambda\bar{\Lambda}}=\meio\frac{1}{\Phi\bar{\Phi}+\Lambda\bar{\Lambda}}\\
\mr{g}_{ij}=0\mbox{, nos demais casos} 
\end{array}\right.
\eeq
onde para realisar os c\'alculos \'e bem mais f\'acil escrever $g$ da forma
\beq
g=(\mr{det} M)^{-\frac{1}{2}-\frac{i}{2}}M,\mbox{ onde, } M=\left(\begin{array}{rr} \Lambda&\bar{\Phi}\\-\Phi&\bar{\Lambda} \end{array}\right).
\eeq

Podemos definir um ``elemento de linha'' neste espa\c{c}o como:
\beq\label{eq.metricaN=2}
ds^2=\frac{1}{\Phi\bar{\Phi}+\Lambda\bar{\Lambda}}(d\Phi d\bar{\Phi}+d\Lambda d\bar{\Lambda}).
\eeq

\'E importante observar que esta m\'etrica \'e invariante por transforma\c{c}\~oes $U(2)$. Para constatar, escrevamos $z^a=\left(\Phi \atop \Lambda\right)$ e $\bar{z}^{\bar{a}}=\left(\bar{\Phi} \atop \bar{\Lambda}\right)$. Assim, uma transforma\c{c}\~ao de $U(2)$ \'e implementada via:
\beq
z'^a=U^a{}_b z^a;\quad \bar{z}'^{\bar{a}}=\bar{U}^{\bar{a}}{}_{\bar{b}} \bar{z}^{\bar{a}},
\eeq
onde $\bar{U}=U^\dagger=U^{-1}.$ Podemos reescrever o termo $\Phi\bar{\Phi}+\Lambda\bar{\Lambda}$ como sendo $z^a\bar{z}^{\bar{a}}\delta_{a\bar{a}}$, onde o s\'{\i}mbolo delta \'e definido atrav\'es de $\delta_{\Phi\bar{\Phi}}=\delta_{\Lambda\bar{\Lambda}}=1$ e $0$ nos outros casos. Assim, fica claro que 
\beq
z'^a\bar{z}'^{\bar{a}}\delta_{a\bar{a}}=U^a{}_b\bar{U}^{\bar{a}}{}_{\bar{b}}\delta_{a\bar{a}}z^b\bar{z}^{\bar{b}}=z^b\bar{z}^{\bar{b}}\delta_{b\bar{b}}.
\eeq
Logo, a m\'etrica ser\'a invariante por estas transforma\c{c}\~oes
\beq
\mr{g}'_{a\bar{a}}=\frac{1}{\Phi\bar{\Phi}+\Lambda\bar{\Lambda}}U_a{}^b\bar{U}_{\bar{a}}{}^{\bar{b}}\delta_{b\bar{b}}=\frac{1}{\Phi\bar{\Phi}+\Lambda\bar{\Lambda}}\delta_{a\bar{a}}.
\eeq
Isto j\'a seria de se esperar pois estamos lidando com a variedade de $SU(2)$.

Desta forma podemos classificar g como sendo do tipo K\"ahler, e assim pode ser escrita em termos de um potencial, como em (\ref{eq.kahlermetrica}). 
Como definido em \cite{Rocek}, o potencial de K\"ahler desta m\'etrica, em termos do conte\'udo de campo apresentado, escreve-se como:

\beq
K=-\left.\int\frac{dx}{x}\ln(1+x)\right|_{\frac{\Lambda\bar{\Lambda}}{\Phi\bar{\Phi}}}+\ln{\Phi}\ln{\bar{\Phi}}.
\eeq

Este termo \'e chamado de ``potencial de K\"ahler torcido'', como em \cite{Jack}. Esta nomenclatura adv\'em do fato de que este potencial n\~ao determina a m\'etrica da variedade segundo as Eq.~(\ref{eq.kahlermetrica}). Calculando-se as derivadas segundas com rela\c{c}\~ao aos campos da a\c{c}\~ao, teremos:

\beq
\begin{array}{l}\DS
\frac{\del^2K}{\del\Phi\del\bar{\Phi}}=\frac{1}{(\Phi\bar{\Phi}+\Lambda\bar{\Lambda})},\quad\frac{\del^2K}{\del\Phi\del\Lambda}=\frac{1}{(\Phi\bar{\Phi}+\Lambda\bar{\Lambda})}\frac{\bar{\Lambda}}{\Phi}, \\
\DS\frac{\del^2K}{\del\Phi\del\bar{\Lambda}}=\frac{1}{(\Phi\bar{\Phi}+\Lambda\bar{\Lambda})}\frac{\Lambda}{\Phi},\quad\frac{\del^2K}{\del\bar{\Phi}\del\Lambda}=\frac{1}{(\Phi\bar{\Phi}+\Lambda\bar{\Lambda})}\frac{\bar{\Lambda}}{\bar{\Phi}}, \\
\DS\frac{\del^2K}{\del\bar{\Phi}\del\bar{\Lambda}}=\frac{1}{(\Phi\bar{\Phi}+\Lambda\bar{\Lambda})}\frac{\Lambda}{\bar{\Phi}},\quad\frac{\del^2K}{\del\Lambda\del\bar{\Lambda}}=-\frac{1}{(\Phi\bar{\Phi}+\Lambda\bar{\Lambda})}.
\end{array}
\eeq

Observe que o primeiro e o \'ultimo termos fazem parte da m\'etrica (\ref{eq.metricaN=2}), por\'em h\'a uma troca do sinal do termo $K_{\Lambda \bar{\Lambda}}$. No entanto, devido \`a propriedades dos campos quirais torcidos, este sinal trocado se compensara na a\c c\~ao e obteremos a m\'etrica correta. Os demais termos compor\~ao a tor\c{c}\~ao. Substituindo os valores acima na a\c{c}\~ao (\ref{eq.Kn=12}), onde integraremos em $\theta_2$, teremos a formula\c c\~ao N=1 no modelo sigma N=2:  


\begin{eqnarray}\label{eq.sigmaN=21}
\lefteqn{\s=\int d^2xd^2\theta_1 \frac{1}{d}\left[D^\al_1\phi D_{1\al}\bar{\phi}+D^\al_1\lambda D_{1\al}\bar{\lambda}+\left(-\frac{\bar{\lambda}}{2\phi}D_{1\al}\phi  D_{1\be}\lambda\right.\right.}\nonumber\\
&&\left.\left.+\frac{\bar{\lambda}}{2\bar{\phi}}D_{1\al}\bar{\phi} D_{1\be}\lambda -\frac{\lambda}{2\phi}D_{1\al}\phi D_{1\be}\bar{\lambda}-\frac{\lambda}{2\bar{\phi}}D_{1\al}\bar{\phi} D_{1\be}\bar{\lambda}\right)\tilde{\delta}^{\al\be}\right].
\end{eqnarray}

Para que esta a\c c\~ao fique semelhante \`a a\c c\~ao (\ref{eq.sigmasusy1}), agrupemos os campos $\phi$, $\bar{\phi}$, $\lambda$ e $\bar{\lambda}$ em um vetor 
\beq
\V\;^i=(\phi, \bar{\phi}, \lambda, \bar{\lambda})
\eeq
e agrupando os coeficientes sim\'etricos e antissim\'etricos, podemos ent\~ao reescrever 
\beq
\s=\int d^2xd^2\theta_1 \left[\mr{g}_{ij}\epsilon^{\al\be}+\mr{b}_{ij}\tilde{\delta}^{\al\be}\right]D_{1\be}\V\;^iD_{1\al}\V\;^j,
\eeq
onde
\begin{eqnarray}\DS
\mr{g}_{ij}&=&\left\{\begin{array}{l}\DS 
\mr{g}_{\phi\bar{\phi}}=\mr{g}_{\lambda\bar{\lambda}}=\frac{1}{2d}\\
\DS0,\qquad\mbox{nos outros casos} 
\end{array}\right.\\
\mr{b}_{ij}&=&\left\{\begin{array}{l}\label{eq.delb}\DS
\mr{b}_{\phi\lambda}=\frac{-\bar{\lambda}}{4d\phi},\qquad \mr{b}_{\bar{\phi}\lambda}=\frac{\bar{\lambda}}{4d\bar{\phi}}\\
\DS\mr{b}_{\phi\bar{\lambda}}=\frac{\lambda}{4d\phi},\qquad \mr{b}_{\bar{\phi}\bar{\lambda}}=\frac{-\lambda}{4d\bar{\phi}}\\
\DS\mr{b}_{ij}=-\mr{b}_{ji}\\
\DS0\mbox{, nos outros casos}
\end{array} \right. 
\end{eqnarray}
 
Se considerarmos que $\V\;^i$ seja o vetor posi\c c\~ao de uma variedade complexa, g e b far\~ao o papel da m\'etrica e da tor\c c\~ao, nesta variedade.

\subsection{O termo de Wess-Zumino-Witten}


Sabemos que o termo de WZW est\'a associado com a tor\c{c}\~ao da variedade de grupo atrav\'es da Eq.~(\ref{eq.WZWH}). Ent\~ao podemos nos perguntar qual a rela\c{c}\~ao entre o campo antissim\'etrico, b,  da a\c{c}\~ao (\ref{eq.sigmaN=21}) e a tor\c{c}\~ao da variedade de grupo deste modelo. Como visto na se\c{c}\~ao \ref{sec.tetrada}, a tor\c{c}\~ao de uma variedade de grupo \'e bem definida atrav\'es da Eq.(\ref{eq.Tg}), ent\~ao, teremos o tensor de Tor\c{c}\~ao definido como:

\beq \label{eq.WZWsusy}   
T_{ijk}=2 \tr \left( g^{-1}\del_{[i} g g^{-1}\del_j g g^{-1}\del_{k]} g \delta^{\al\be}\right).
\eeq

Podemos expressar a tor\c c\~ao em termos de campos N=2, calculando explicitamente suas componentes atrav\'es da defini\c c\~ao do elemento de grupo, Eq.~(\ref{eq.elementoN=2}): 



\beq \label{eq.compT}
\left\{\begin{array}{rcl} 
T_{\Phi\bar{\Phi}\Lambda}&=&\frac{-1}{d^2}\bar{\Lambda}\\
T_{\Phi\bar{\Phi}\bar{\Lambda}}&=&\frac{1}{d^2}\Lambda\\
T_{\Phi\Lambda\bar{\Lambda}}&=&\frac{-1}{d^2}\bar{\Phi}\\
T_{\bar{\Phi}\Lambda\bar{\Lambda}}&=&\frac{1}{d^2}\Phi
\end{array}\right.
\eeq

Vejamos que podemos relacionar as componentes de $T$ com as componentes de b, definidas em (\ref{eq.delb}), atrav\'es de:
\begin{eqnarray}
\del_{[\bar{\phi}}\mr{b}_{\phi\lambda]}&=&\frac{\bar{\lambda}}{6d^2}\\
\del_{[\bar{\lambda}}\mr{b}_{\phi\lambda]}&=&-\frac{\bar{\phi}}{6d^2}\\
\del_{[\bar{\phi}}\mr{b}_{\phi\bar{\lambda}]}&=&-\frac{\lambda}{6d^2}\\
\del_{[\bar{\lambda}}\mr{b}_{\bar{\phi}\lambda]}&=&\frac{\phi}{6d^2}
\end{eqnarray}

Ent\~ao vemos que o termos de WZW deste modelo \'e proporcional a tor\c{c}\~ao da variedade de grupo de $SU(2)\otimes U(1)$. 

Analisando a quest\~ao topol\'ogica apresentada na se\c c\~ao \ref{sec.top}, veremos que aqui tamb\'em teremos a presen\c ca do {\it winding number}. Para isso, basta observar\footnote{Podemos ver que isto \'e verdade lembrando que $U(2)=S^3\otimes S^1$, que $\pi_2(S^3)=\pi_2(S^1)=0$, e que $\pi_3(S^3)=\Z$ como explicado em \cite{Actor}.} que 
\beq
\pi_2(U(2))=0 \mbox{, e } \pi_3(U(2))=\Z.
\eeq

Recapitulando, t\'{\i}nhamos uma representa\c c\~ao N=2 do grupo $SU(2)\otimes U(1)$ cuja m\'etrica pode ser escreita em termos de um potencial $K$. Utilizando o fato de $K$ ser composto por campos quirais e quirais torcidos N=2, pudemos reescrever o modelo em termos de campos N=1. Nesta forma, pudemos compar\'a-lo com a forma geral do modelo \sig N=1. A partir da\'{\i} relacionamos o termo   antissim\'etrico com a tor\c c\~ao  da variedade de grupo. 

C\'alculos da fun\c c\~ao beta deste tipo de modelo podem ser encontradas em \cite{Jack}, onde apenas para c\'alculos de quatro loops teremos a fun\c c\~ao beta diferente do tensor de Ricci.


\chapter*{Conclus\~ao e Perspectivas}
\addcontentsline{toc}{chapter}{Conclus\~ao e Perspectivas}
\markright{Conclus\~ao e Perspectivas}

Neste trabalho apresentamos alguns aspectos importantes do modelo sigma n\~ao linear e do termo de WZW. Analisando o modelo em duas dimens\~oes, observamos que o termo de WZW relaciona-se com a tor\c c\~ao da variedade de grupo, permitindo assim uma descri\c c\~ao sucinta das equa\c c\~oes de movimento quando utilizamos a conex\~ao de Christofell acrescida da tor\c c\~ao. Outro ponto relevante \'e a quest\~ao topol\'ogica do termo de WZW, pois este n\~ao depende da m\'etrica. A rela\c c\~ao com variedades de grupo fica evidenciada quando descrevemos o modelo via elementos de um grupo {\got G}.

Em duas dimens\~oes, vimos tamb\'em que o modelo N=1 n\~ao imp\~oe nenhuma condi\c c\~ao sobre a variedade alvo, podendo esta ser exatamente igual \`a mesma do caso bos\^onico. O caso N=2 se mostra mais interessante a come\c car pelo fato de sua din\^amica depender da estrutura cinem\'atica dos campos da a\c c\~ao. Este modelo imp\~oe restri\c c\~oes sobre a variedade alvo, onde esta deve ser twisted K\"ahler. Apesar de n\~ao ter ficado expl\'{\i}cito, a presen\c ca dos campos quirais modificam a estrutura da variedade K\" ahler que obter\'{\i}amos se apenas tiv\'essemos o modelo sigma principal. Deve-se aqui comentar que, em quatro dimens\~oes, j\'a no caso N=1 teremos a variedade alvo como sendo K\"ahler, e em seis dimens\~oes, este ter\'a de ser hiper-K\"ahler. Modelos sigma supersim\'etricos em mais dimens\~oes n\~ao s\~ao poss\'iveis.

Um ponto que ficou de lado neste trabalho foi a quest\~ao da invari\^ancia conforme do modelo sigma. Apresentamos alguns argumentos que refor\c cam a id\'eia de que para termos invari\^ancia conforme, devemos ter que o termo de WZW seja exatamente a tor\c c\~ao da variedade de grupo. Para o caso sem tor\c c\~ao, ficou claro que a variedade alvo tem que obedecer as equa\c c\~oes de Einstein. Pode-se ver na refer\^encias (ex. \cite{Jack}) que o mesmo se repete para o modelo completo, por\'em utilisando a tor\c c\~ao. 

As perspectivas de continua\c c\~ao deste estudo s\~ao bastante variadas. Entre os assunto que infelizmente n\~ao foram abordados por falta de tempo, poder\'{\i}amos cita a rela\c c\~ao entre as supersimetrias e o fato de a fun\c c\~ao beta ser nula, propriedade esta que ocorre no caso N=4, mas que n\~ao \'e clara para modelos N=2. Neste t\'opico ainda temos o estudo pormenorizado dos modelos com N=4, onde temos uma estrutura quaterni\^onica. Outro ponto \'e o estudo da rela\c c\~ao entre o termos pricipal, o termo de WZW e a variedade de grupo, esta que depende da representa\c c\~ao dos supercampos, onde no nosso caso vimos que os campos quirais e quirais torcidos implicam em variedade K\"ahler torcida.  

Poderiamos tamb\'em analizar a dualidade entre os modelos sigma compostos por representa\c c\~oes diferentes do mesmo grupo (p.e. uma com campos quirais e outro com campos semi-quirais), estudar as propriedades fora da camada de massa dos modelos. Seria importante estudadar mais a quantiza\c c\~ao destes modelos  e sua rela\c c\~ao com teoria de cordas.

\end{document}